\documentclass[a4paper,onecolumn,11pt]{quantumarticle}
\pdfoutput=1
\usepackage[utf8]{inputenc}
\usepackage[english]{babel}
\usepackage[T1]{fontenc}
\usepackage{amsmath}
\usepackage{hyperref}

\usepackage{tikz}
\usepackage{lipsum}

\usepackage{amssymb}
\usepackage{epsf,float,epsfig}
\usepackage[numbers,sort&compress]{natbib}

\begin{document}

\title{Nested Grover's Algorithm  for Tree Search}

\author{Andreas Wichert}
\affiliation{Department of Computer Science and Engineering,  INESC-ID  \&  Instituto Superior T\'ecnico, University of Lisbon, {2740-122} Porto Salvo, Portugal}
\orcid{0000-0002-2179-4378}
\email{andreas.wichert@tecnico.ulisboa.pt}
\homepage{http://web.ist.utl.pt/andreas.wichert/}
\orcid{0000-0003-0290-4698}

\maketitle

\begin{abstract}
  We investigate optimizing quantum tree search algorithms by employing a nested Grover Algorithm. This approach seeks to enhance results compared to previous Grover-based methods by expanding the tree of partial assignments to a specific depth and conducting a quantum search within the subset of remaining assignments. The study explores the implications and constraints of this approach, providing a foundation for quantum artificial intelligence applications.
Instead of utilizing conventional heuristic functions that are incompatible with quantum tree search, we introduce the partial candidate solution, which indicates a node at a specific depth of the tree. By employing such a function, we define the concatenated oracle, which enables us to decompose the quantum tree search using Grover’s algorithm.
With a branching factor of $2$ and a depth of $m$, the costs of Grover’s algorithm are $O(2^{m/2})$. The concatenated oracle allows us to reduce the cost to $O(m \cdot 2^{m/4})$ for $m$ partial candidate solutions.
\end{abstract}

\section{Introduction}

We investigate the optimization of quantum tree search algorithms by employing a nested Grover Algorithm. We explore the implications of this approach and identify the resulting constraints. Our findings demonstrate the potential for enhanced results compared to previously developed Grover-based methods.
Instead of adopting the hybrid divide-and-conquer methods that are predicated on quantum backtracking and quantum walk as indicated in \cite{Montanaro2018}, \cite{Rennela2923}, we will proceed with the initial conception of a nested version of Grover’s nested search, as described in \cite{Cref2000}.
In a nested version of Grover search, the complete tree of partial assignments is expanded to a specific depth, and then a quantum search is conducted within the subset of partial assignments that have not yet been eliminated.

It is claimed that, for certain reasonable distributions of problems, the average complexity of the original nested version of Grover search  \cite{Cref2000} will be less than that obtained from Grover search. This claim is supported by references \cite{Cref2000} and \cite{Montanaro2018}.
Building upon the initial concept of a nested version of Grover search, we will adopt a combinatorial approach to address the limitations inherent in the original work as described in \cite{Cref2000}.
For simplicity, we assume that the tree is uniform and its size is known beforehand. We will demonstrate our approach on simple examples using Qiskit, see \cite{Qiskit2023}.
The main contributions of this paper are as follows:
\begin{itemize}
\item  We introduce the definition of a partial candidate solution and the contented oracles.
\item  We analyze the constraints of the original Grover’s nested search as introduced in \cite{Cref2000}.
\item  We introduce the iterative approach with a quadratic speedup in comparison to the original Grover’s algorithm.
\item  We investigate the possibilities of dividing the original space into two disentanglement sunspaces and the resulting consequences.
\item  We introduce the concept of performing a permutation on a subspace and the corresponding constraint.
\end{itemize}
Our approach imposes specific constraints that elucidate the concepts of partial candidate solutions within the context of quantum tree search.
This work provides the foundation for quantum artificial intelligence applications, including quantum problem-solving and quantum production systems.

\section{Quantum Tree Search}

\begin{figure}[htb]
\begin{center}
\leavevmode
\epsfysize7cm
\epsffile{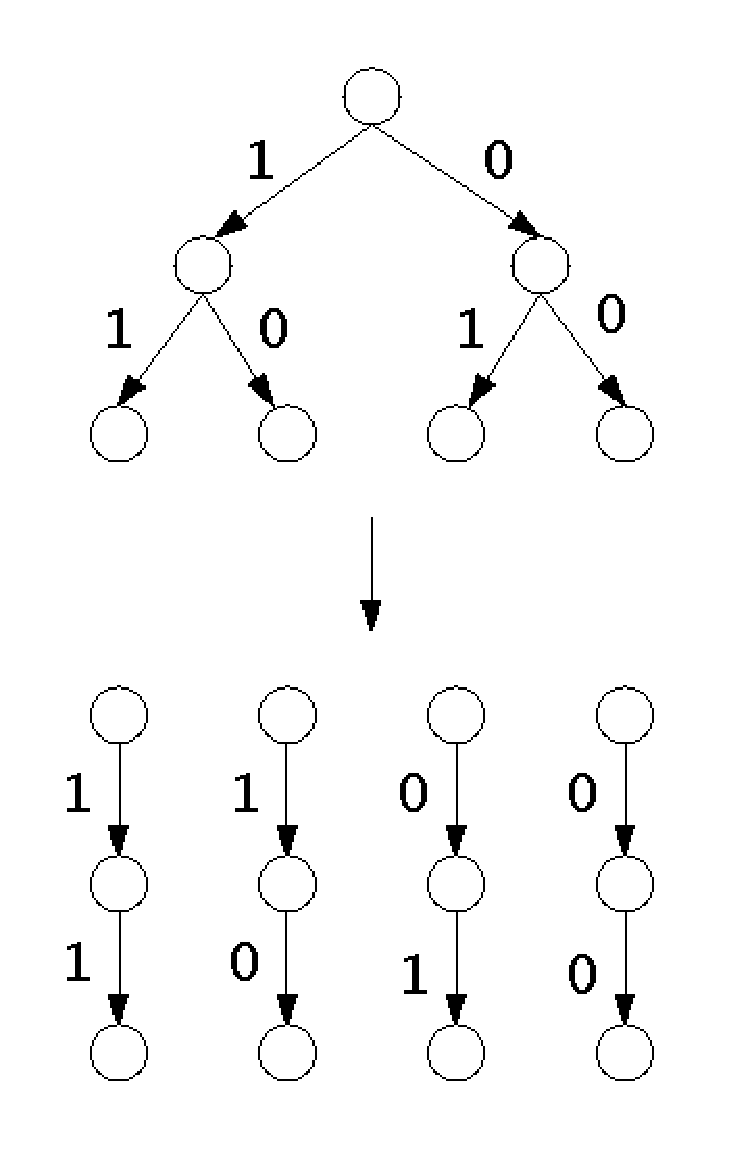}
\end{center}
\caption{ Search tree for $B=2$ and depth  $m=2$. Each question can be represented by a bit. Each binary number ($11$, $10$, $01$, $00$) represents a path descriptor $m$ from the root to the leaf.} \label{bt.eps}
\end{figure}

\begin{figure}[htb]
\begin{center}
\leavevmode
\epsfysize3.4cm
\epsffile{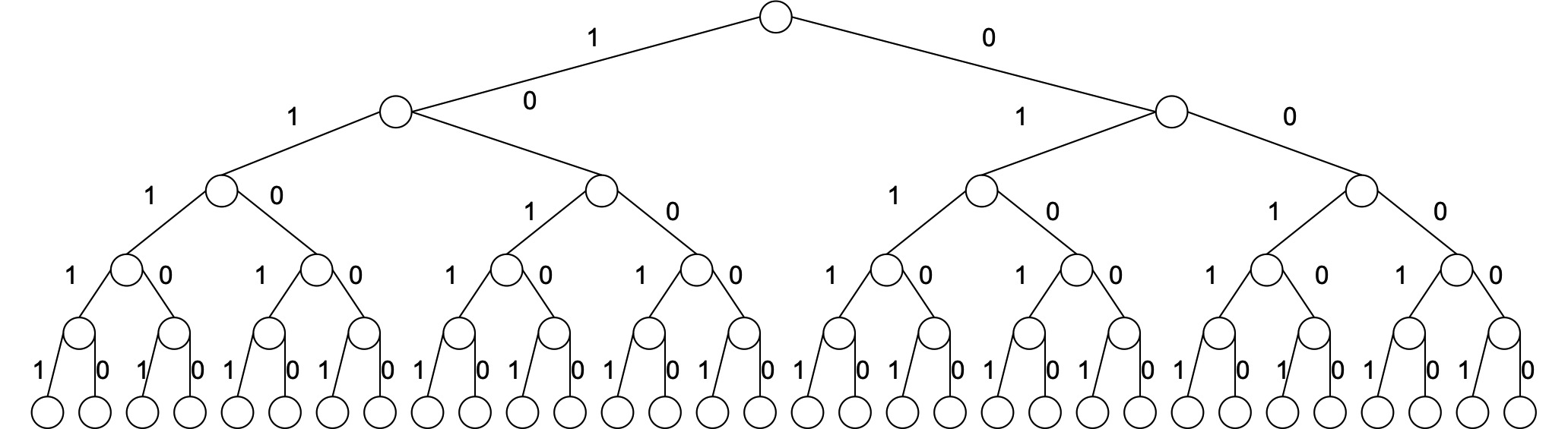}
\end{center}
\caption{Search tree for $B=2$ and depth  $m=5$. Each question can be represented by a bit.}
\label{Tree_2_3.eps}
\end{figure}

In this section, we outline the application of Grover’s algorithm to the tree search problem.
Nodes and edges in a search tree represent states and transitions between states \cite{Wichert2020b}. The initial state is the root, and from each state, either $B$ states can be reached or it’s a leaf.
From leaf no other state can be reached. $B$ represents the branching factor of a node, indicating the number of possible choices. A leaf can either be the goal of the computation or an impasse when there are no valid transitions to a succeeding state.
Each node except the root has a unique parent node, which is called the parent. Every node and its parent are connected by an edge. Each parent has $B$ children. If $B=2$ and the depth of the tree is $m$, each of the $m$ questions has a yes/no reply and can be represented by a bit (as shown in Figure \ref{bt.eps}). The $m$ answers are collectively represented by a binary number of length $m$.  
There are $n=2^m=B^m$ possible binary numbers of length $m$. Each binary number represents a path from the root to a leaf. For each goal, a specific binary number indicates the solution. This binary number is called a path descriptor, since it describes a path from the root of the tree to the leaf, see Figure \ref{Tree_2_3.eps} with  $B=2$ and depth  $m=5$.
For a constant branching factor $B>2$, each question has $B$ possible answers. The $m$ answers can be represented by a base-$B$ number with $m$ digits.
In a quantum computation, we can simultaneously represent all possible path descriptors. There’s one path descriptor for each leaf off the tree.
Using Grover’s algorithm, we search through all possible paths and verify whether each path leads to the goal state. This type of procedure is known as a quantum tree search \cite{tarrataca2010}. For $n=B^m$ possible paths, the costs are (approximately) $\sqrt{n}=B^\frac{m}{2}$ corresponding to the reduced branching factor $b_q=\sqrt{B}$.

\subsection{Grover's Algorithm}

For a function $o(x)$
\begin{equation}
  o_{\xi}(x)= \left\{
  \begin{array}{l} 
1~~~ if~~~x=\xi  \\
 0~~~else
 \end{array}  \right.
 \end{equation}
 we seek to identify the value of $x$ representing a path descriptor for which $o(x)=1$, where $x=\xi$, the path descriptor leading to the solution. This task is equivalent to a decision problem with a binary answer: $1$ indicating a successful search and $0$ indicating an unsuccessful search. The instance $x$ is the input to the problem.
 
Grover’s amplification algorithm implements exhaustive search in $O(\sqrt{n})$ steps in an $n$-dimensional Hilbert space   \cite{grover1996},  \cite{grover1997},  \cite{grover1998a},  \cite{grover1998b},  \cite{grover1996},  \cite{grover1996}. 
This algorithm is derived from the Householder reflection of the state $|x\rangle$ of $m$ qubits, where $n=2^m$. Grover’s amplification algorithm is optimal for exhaustive search, as demonstrated by the lower bound $\Omega(\sqrt{n})$ established by  \cite{aharonov1999}.
Grover's amplification algorithm is optimal, one can prove that a better algorithm cannot exist \cite{bennett1997},  \cite{boyer1998}. It follows  that using a quantum computer $NP-complete$ problems remain $NP-complete$.   Grover’s amplification algorithm provides a quadratic speedup over classical computers, which would require $n$ steps to solve the problem. 

For $k$ solutions and  more than two qubits ($m >2$, $n=2^m$) the number $r$ of iterations  is the largest integer 
\begin{equation}
r=\left\lfloor \frac{\pi}{4} \cdot \sqrt{\frac{2^m}{k}} \right\rfloor.
\label{eq:rotation}
 \end{equation}
The value of $r$ depends on the relation of $n$ versus $k$. For $n=4$ and $k=1$ we need only one rotation, we need as well only one rotation for
\[  \frac{n}{4}=k \]
to find \textbf{one} of the $k$ solutions. 
The iterative amplification algorithm requires the value of $k$ in order to determine the number of iterations. We can determine the value of $k$ by the quantum counting algorithm  \cite{brassard1998}, \cite{brassard2000},  \cite{kaye2007a}.

Uniform distributions are essential for the Grover's algorithms. If the distribution is non-uniform, the algorithms may not function properly or require adaptation which leads to the same complexity  $O(\sqrt{n})$  \cite{Ventura1988, Ventura2000, Cref2000, tay2010}. Consequently, we cannot simply mark potential solutions by assigning higher amplitude values to speed up the search. For instance, an adapted algorithm for sparse distributions where the majority of amplitudes are zero must generate a uniform distribution for the unmarked sets with the resulting complexity  $O(\sqrt{n})$  \cite{Ventura1988, Ventura2000, tay2010}.

In our analysis in relation to the $O$ notation  $O(\sqrt{\frac{n}{k}})$
and for simplification we will assume the approximate complexity of Grover's algorithm as
\begin{equation}
\sqrt{\frac{2^m}{k}} =\sqrt{\frac{n}{k}}. 
 \end{equation}
 instead of using the accurate Equation \ref{eq:rotation}.

\section{Nested Grover's Search}

Is it possible to integrate nested Grover’s search techniques into the quantum tree search algorithm to enhance the search process?
Can the nested Grover’s search be based on heuristic functions? A heuristic function $\mu(y)$ estimates the cheapest cost from a node to the goal. The local path descriptor $y$ describes the path from the root to the node. Therefore, each node in the tree can be described by a local path descriptor. However, we cannot employ a heuristic function to generate a non-uniform distribution, as it does not confer any advantage in Grover’s search. Furthermore, we cannot compare branches represented by distinct path descriptors. This is also the reason why the min-max algorithm cannot be applied to quantum tree search in games. The reason for this is that distinct branches correspond to distinct superpositions, and these superpositions interact solely through interference.

\subsection{Partial Candidate Solution}

Within the context of quantum tree search, instead of employing a heuristic function $\mu(y)$, we will utilize the function $h(y)$, which represents a partial candidate solution and $y$ denotes a path from the root to the corresponding node. In contrast to the heuristic function $\mu(y)$, which can mark multiple nodes, $h(y)$ should exclusively mark a single partial solution that includes the node indicated by the path $y$.

\subsection{Decomposition}

\begin{figure}[htb]
\begin{center}
\leavevmode
\epsfysize3.4cm
\epsffile{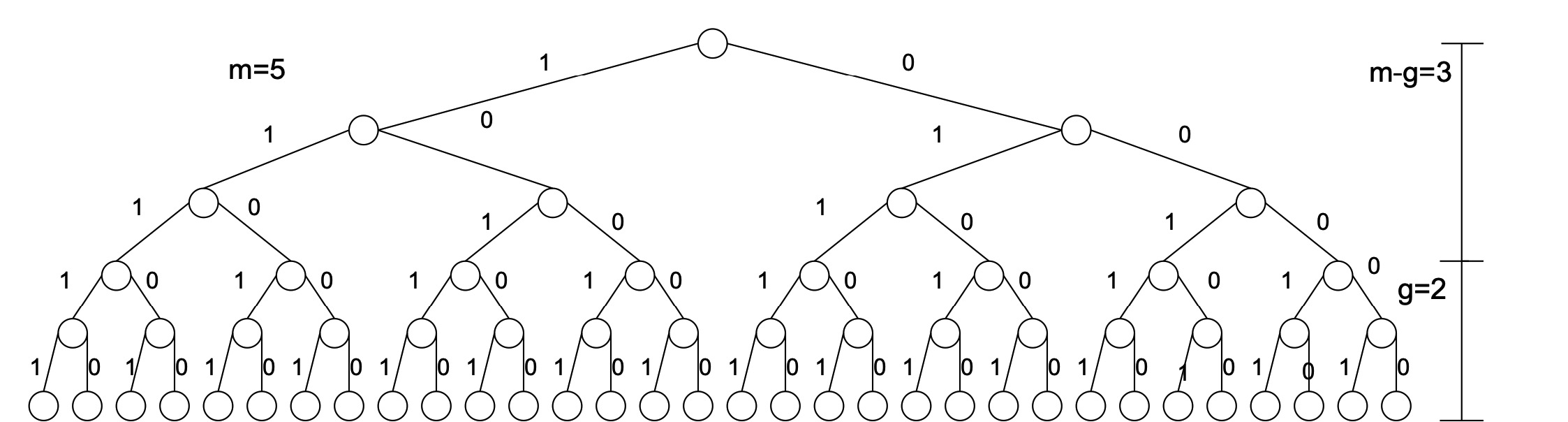}
\end{center}
\caption{The  decomposition  of the  Hilbert space $\mathcal{H}$ of dimension $n=2^m$ into
subspaces  $\mathcal{L}$ and  $\mathcal{U}$ of dimension $2^g$ and $2^{m-g}$. In our example $m=5$ and $g=2$. }
\label{Tree_2_3_names.eps}
\end{figure}

\begin{figure}[htb]
\begin{center}
\leavevmode
\epsfysize5cm
\epsffile{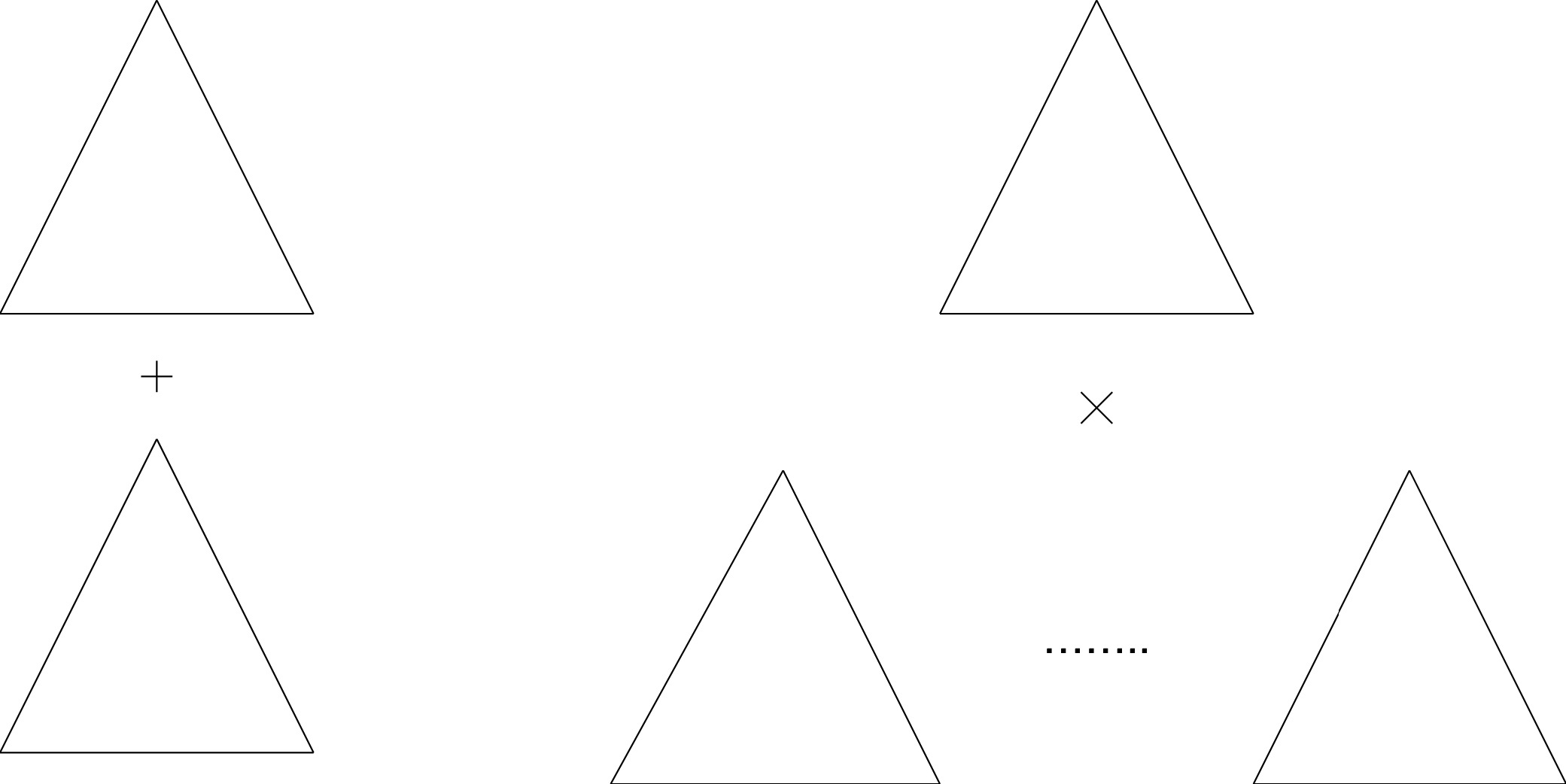}
\end{center} 
\caption{The concept of cost savings is predicated on the inequality that decomposes the Hilbert space $\mathcal{H}$ of dimension $n=2^m$ into subspaces $\mathcal{H}=\mathcal{U}   \otimes \mathcal{L}$ of dimensions $2^g$ and $2^{m-g}$, respectively. On the left side, the sum of the costs of Grover’s algorithm, with leaves for addition resulting in $\sqrt{2^g} + \sqrt{2^{m-g}}$, is represented. On the right side, the original costs of Grover’s algorithm are $\sqrt{2^m}=2^g \cdot 2^{m-g}$, corresponding to the multiplication operation. }
\label{Add_mul.eps}
\end{figure}

The idea of saving the costs is based on the inequality, that represents the  decomposition  of the  Hilbert space $\mathcal{H}$ of dimension $n=2^m$ into
subspaces  $\mathcal{L}$ and  $\mathcal{U}$ of dimension $2^g$ and $2^{m-g}$ , see Figure \ref{Tree_2_3_names.eps}.
The costs of Grover's algorithm for the quantum tree search algorithm are
\[\sqrt{2^m}=\sqrt{2^g} \cdot \sqrt{2^{m-g}}.\]
However, the costs associated with Grover’s algorithm for the quantum tree of each subspace would be significantly reduced to
\[  \sqrt{2^g} + \sqrt{2^{m-g}} \]
 (see Figure \ref{Add_mul.eps}) with
\begin{equation}
 \sqrt{2^g} + \sqrt{2^{m-g}} < \sqrt{2^g} \cdot \sqrt{2^{m-g}},
 \end{equation} 
since
 \[
 1  <  \sqrt{2^g} \cdot \sqrt{2^{m-g}} - \sqrt{2^g} + \sqrt{2^{m-g}} +1
 \]
  \[
 1  <   (\sqrt{2^g} -1)\cdot (1- \sqrt{2^{m-g}}) 
 \]
 because
 \[
1 < (\sqrt{2^g} -1),~~ 1 <    (1- \sqrt{2^{m-g}}).
 \]

\subsection{Nested Grover's Search and Entanglement}

\begin{figure}
\leavevmode
\parbox[b]{8cm}{ (a) \epsfxsize13cm\epsffile{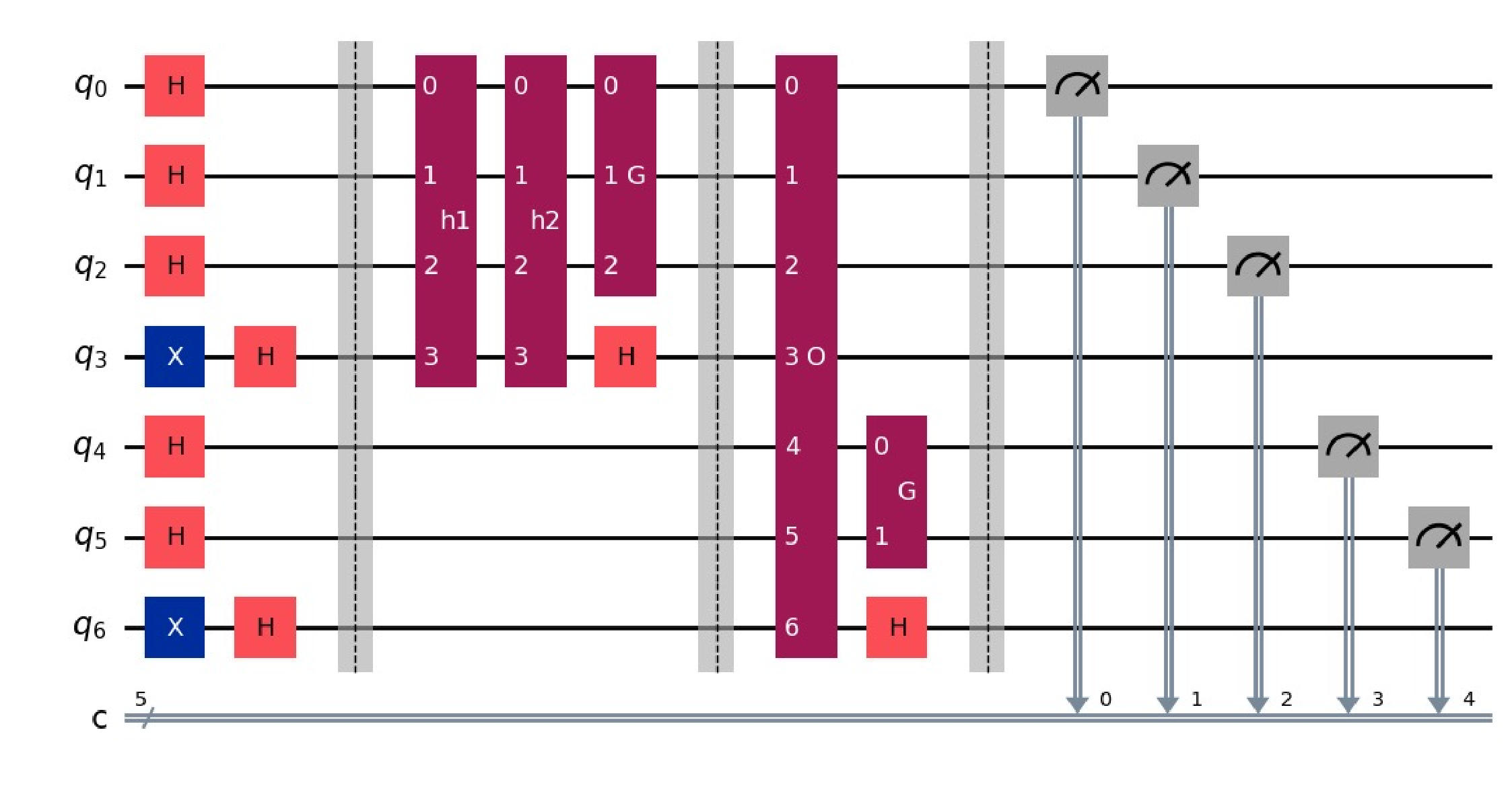}}
\begin{center}
\parbox[b]{7cm}{ (b) \epsfxsize7cm\epsffile{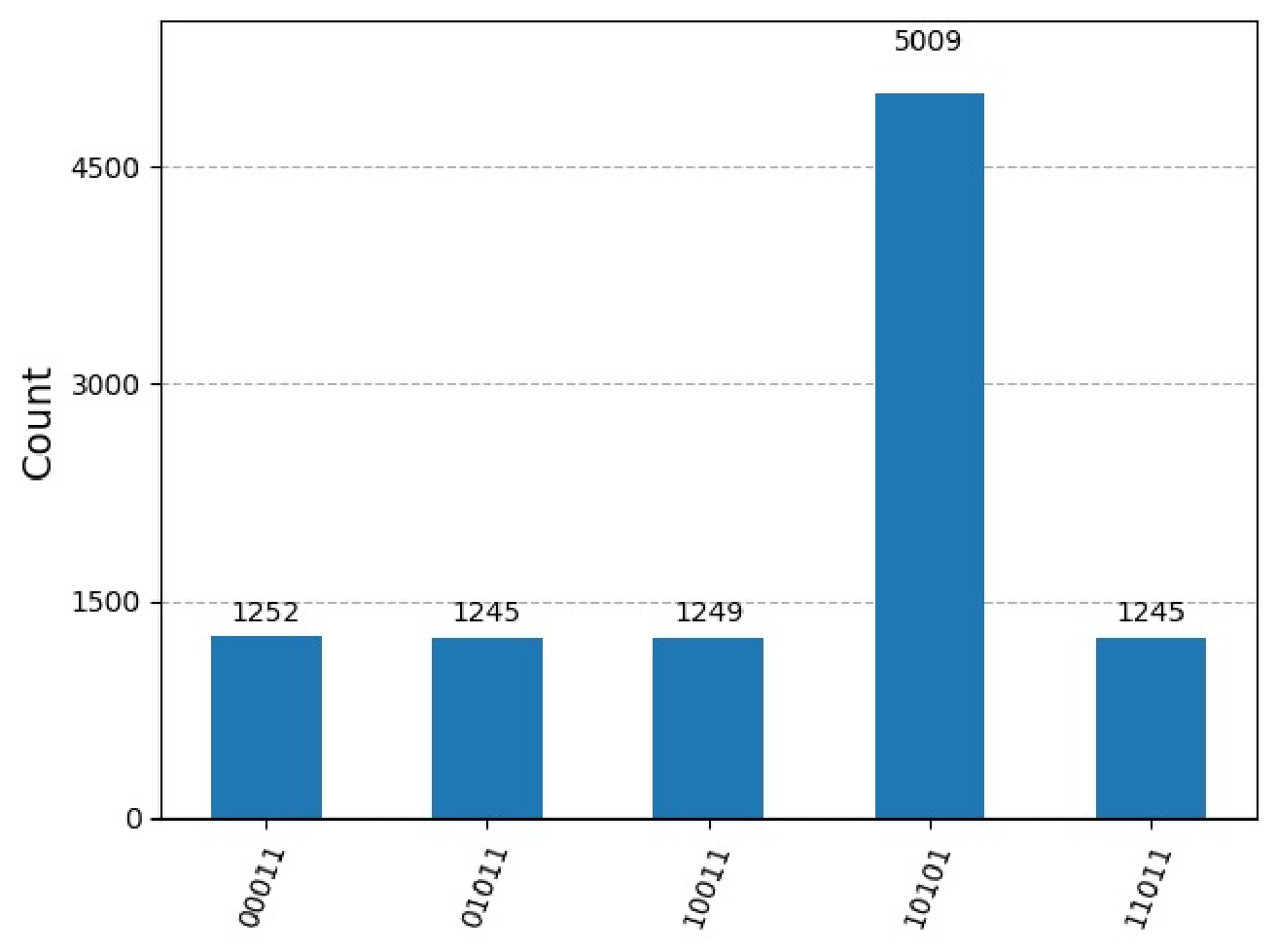}}
\end{center}
\caption{(a) Circuit using Grover's amplification representing oracles $h^{(1)}(y)=\neg x_3 \wedge x_2   \wedge  x_1$, $h^{(2)}(y)=x_3 \wedge \neg x_2   \wedge  x_1$ for lower qubits and the global oracle $ o(x)= x_5 \wedge \neg x_4  \wedge    x_3 \wedge \neg x_2   \wedge  x_1$.  We we will employ the little-endian notation as used in Qiskit, see \cite{Qiskit2023}. (b) Due to the entanglement of the lower qubits with the upper qubits, the solution indicated by $10101$ is indicated only approximately by $50\%$. } 
\label{A_basic_2.eps}
\end{figure}

Following alternatively the idea of Grover  that should speed up the search on  Hilbert space $\mathcal{H}$ of dimension $n=2^m$  a set $\aleph$  of $v$ partial candidate solutions
\[
  h^{(k)}(y)~~with~ k \in \{ 1, 2,\cdots , v   \}
\]
is defined  that acts on $g$ lower qubits with
\begin{equation}
  h^{(k)}(y)= \left\{
  \begin{array}{l} 
1~~~ if~~~y=h_k \\
 0~~~else
 \end{array}  \right. 
 \end{equation}
 on the subspace $\mathcal{L}$ of dimension $2^g$, with $y$ being a local path descriptor.
 Throughout the paper, we will employ the little-endian notation, which stores the least-significant byte at the smallest address. Qubits are represented from the most significant bit (MSB) on the left to the least significant bit (LSB) on the right (little-endian), similar to binary numbers or bit-string representation on classical computers. 
We illustrate the fundamental concept of nested Grover search through a simple SAT example, characterized by $v=2$, $g=3$, and $m=5$.
\[ 
 h^{(1)}(y)=\neg x_3 \wedge x_2   \wedge  x_1,~~ h^{(2)}(y)=x_3 \wedge \neg x_2,   \wedge  x_1~~~y=x_3, x_2, x_1;~~ x_k \in {0,1},
 \]
 \[
 o(x)= x_5 \wedge \neg x_4  \wedge    x_3 \wedge \neg x_2   \wedge  x_1,~~~x=x_5, x_4, \cdots, x_1;~~~x_k \in \{0,1\},
\]
 as indicated by the circuit in Figure \ref{A_basic_2.eps} (a) with the  results of measurement shown in  Figure \ref{A_basic_2.eps} (b). 
 In the initial step, we determine the set $\aleph$ using Grover’s algorithm on the subspace $\mathcal{L}$. In the subsequent step, we anticipate utilizing this set to expedite the search for the solution indicated by the oracle $o_{\xi}(x)$, which operates on $m$ qubits. Grover’s amplification is applied to the subspace $\mathcal{U}$ defined by the upper $m-g$ qubits. However, due to entanglement, we are unable to employ Grover’s algorithm on the upper subspace $\mathcal{U}$ defined by $m-g$ upper qubits, as both subspaces $\mathcal{L}$ and $\mathcal{U}$ become entangled.
To comprehend the behavior we introduce concatenated oracles.

\subsection{Concatenated Oracles}

\begin{figure}[htb]
\begin{center}
\leavevmode
\epsfysize3.4cm
\epsffile{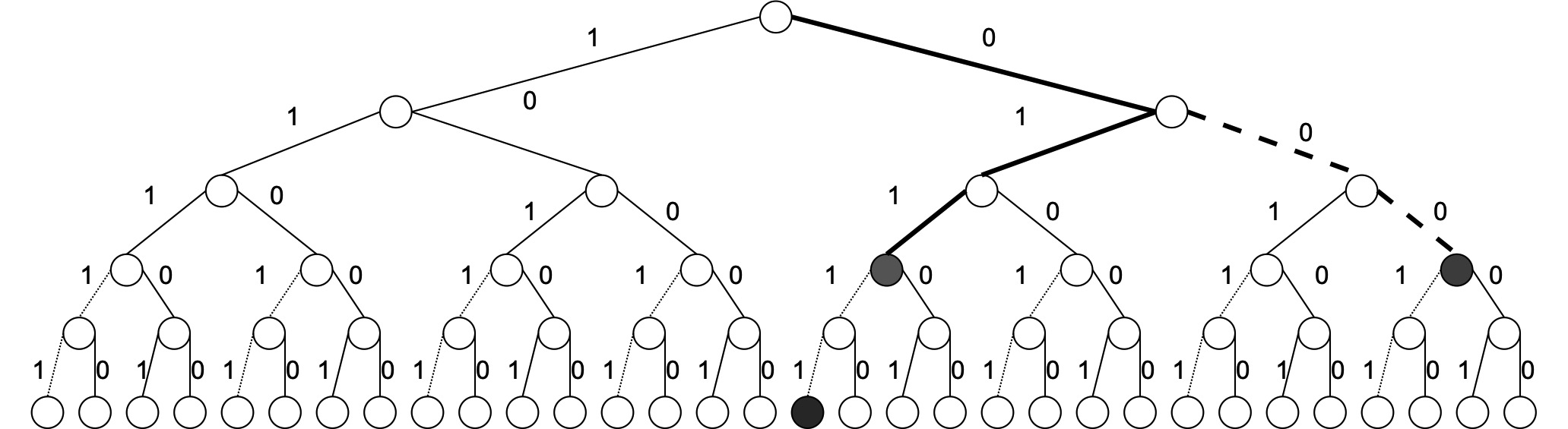}
\end{center}
\caption{Search tree for $B=2$ and depth  $m=5$. With $h_1=011$,  $h_2=000$, $u=11$, $l=011$ and $\xi=01111$.}
\label{Tree_2_3_heuristics.eps}
\end{figure}

Concatenation is an associative operation but not a commutative operation. We use for the operation concatenation the specific symbol $\shortparallel $.
If we assume that $x$ is represented by $m$ bits with the notation
\[
x=x_m, x_{m-1},  x_{m-2}\cdots, x_2, x_1;~~~  x_k \in \{0,1\},
\]
then  concatenation is often simply the placement of the binary numbers next to each other,
\[
(x_m, x_{m-1}, \cdots ,x_{g} ) \shortparallel  (x_{g-1},\cdots,x_1) = x_m, x_{m-1},\cdots, x_2, x_1 
\]
For example a binary number $10$ and $11$ we get
\[
10  \shortparallel 11  = 1011.
\]
We extend the concatenation for the oracle function 
\begin{equation}
 o_{\xi}(x_m, x_{m-1},  x_{m-2}\cdots, x_1 )= u(x_m, x_{m-1}, \cdots, x_{g} ) \shortparallel   l=(x_{g-1},\cdots,x_1).
\end{equation}
Using compact notation,
\[
z= (x_m, x_{m-1},\cdots,x_{g} ),~~y= (x_{g-1},\cdots,x_1),
\]
\[
x= (x_m, x_{m-1},\cdots,x_{g} ) \shortparallel (x_{g-1},\cdots,x_1),
\]
\[
x=z \shortparallel y,
\]
and the oracle function 
\[
  o_{\xi}(x)=u(z)  \shortparallel l(y)
\]
with
\begin{equation}
  o_{\xi}(x)= \left\{
  \begin{array}{l} 
1~~~ if~~~x=\xi  \\
 0~~~else
 \end{array}  \right.
  u(z)= \left\{
  \begin{array}{l} 
1~~~ if~~~z=u  \\
 0~~~else
 \end{array}  \right.
  l(y)= \left\{
  \begin{array}{l} 
1~~~ if~~~y=l  \\
 0~~~else
 \end{array}  \right.
 \end{equation}
 solution being
 \[
\xi = u \shortparallel l,
 \]
 with  $u=u(z)$ being the upper part of the oracle function that acts on the upper qubits  describing the upper subspace $\mathcal{U}$  of the dimension $2^{(m-g)}$ and $ l=l{(y)}$ being the lower part of the oracle function that acts on the lower qubits describing the lower subspace  $\mathcal{L}$ of the dimension $2^g$
with
\[ o(x) = o(z,y)= u(z) \shortparallel l(y).  \]
Note that $l(y)$ and $u(z)$ do not indicate partial candidate solutions; they are exact copies of the oracle function, and it can be challenging to specify them precisely.
Also  $u$ indicates the path descriptor. An example  is represented in Figure \ref{Tree_2_3_heuristics.eps}.
In the case we constraint $y$ values to some values $h_k$ with
\[
y \in \{h_1, h_2, \cdots, h_v \} 
\]
the constraint oracle can be expressed as $l=h_k$ with
\[ o(x) = o(z,h_k)=  u(z) \shortparallel l(h_k).  \]

\subsection{Not Entangled Subspaces}

Imagine we know $u_{\mu}(z)$ that acts on the upper subspace  $\mathcal{U}$ defined by $(m-g)$  lower qubits and  given set $\aleph$  of $t$ partial candidate solutions   that act on the subspace  $\mathcal{L}$ defined by upper $g$ qubits,
\[
  h^{(k)}(y)~~with~ k \in \{ 1, 2,\cdots , v  \}.
\]
If we apply both oracles and use Grover's algorithm to each Hilbert space defined by the subspace  $\mathcal{L}$ and $\mathcal{U}$,  the results are not entangled since both subspaces are not entangled.
In the first step, we determine $\aleph$ using Grover’s algorithm on the Hilbert space defined by the subspace  $\mathcal{L}$. In the second step, we determine $u(z)$ using Grover’s algorithm on the upper qubits of the  upper subspace $\mathcal{U}$ with
\begin{equation}
\mathcal{H}=\mathcal{U}   \otimes \mathcal{L}. 
\end{equation}
If we measure both subspaces, we see that each solution in the lower subspace  $\mathcal{L}$ defines a new combination of the one solution in the upper subspace $\mathcal{U}$. One of the combination could correspond to the global solution of the space  as described by the oracle   $o_{\xi}(x)$, however we do not know if this is the case.

\begin{figure}
\leavevmode
\parbox[b]{8cm}{ (a) \epsfxsize13cm\epsffile{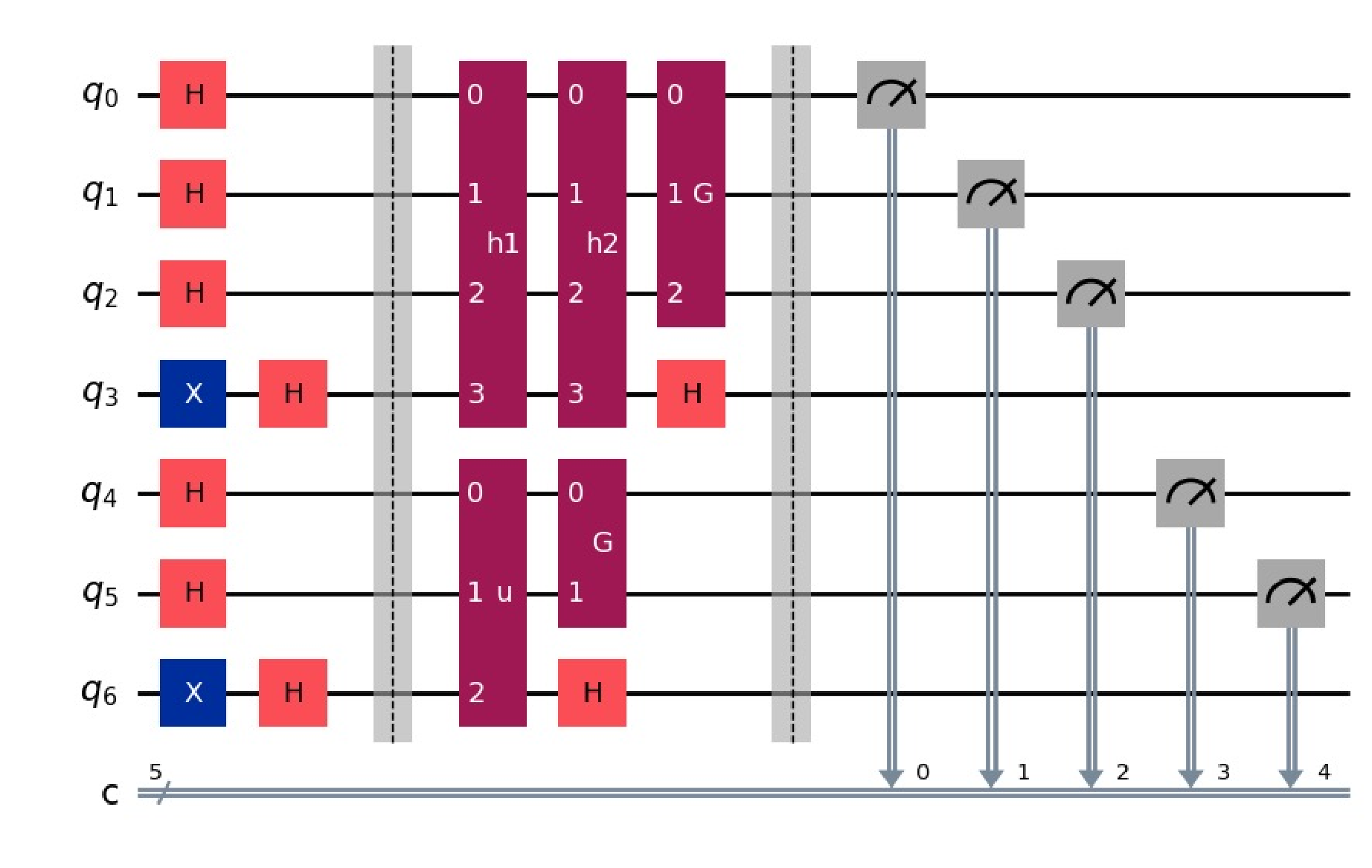}}
\begin{center}
\parbox[b]{7cm}{ (b) \epsfxsize7cm\epsffile{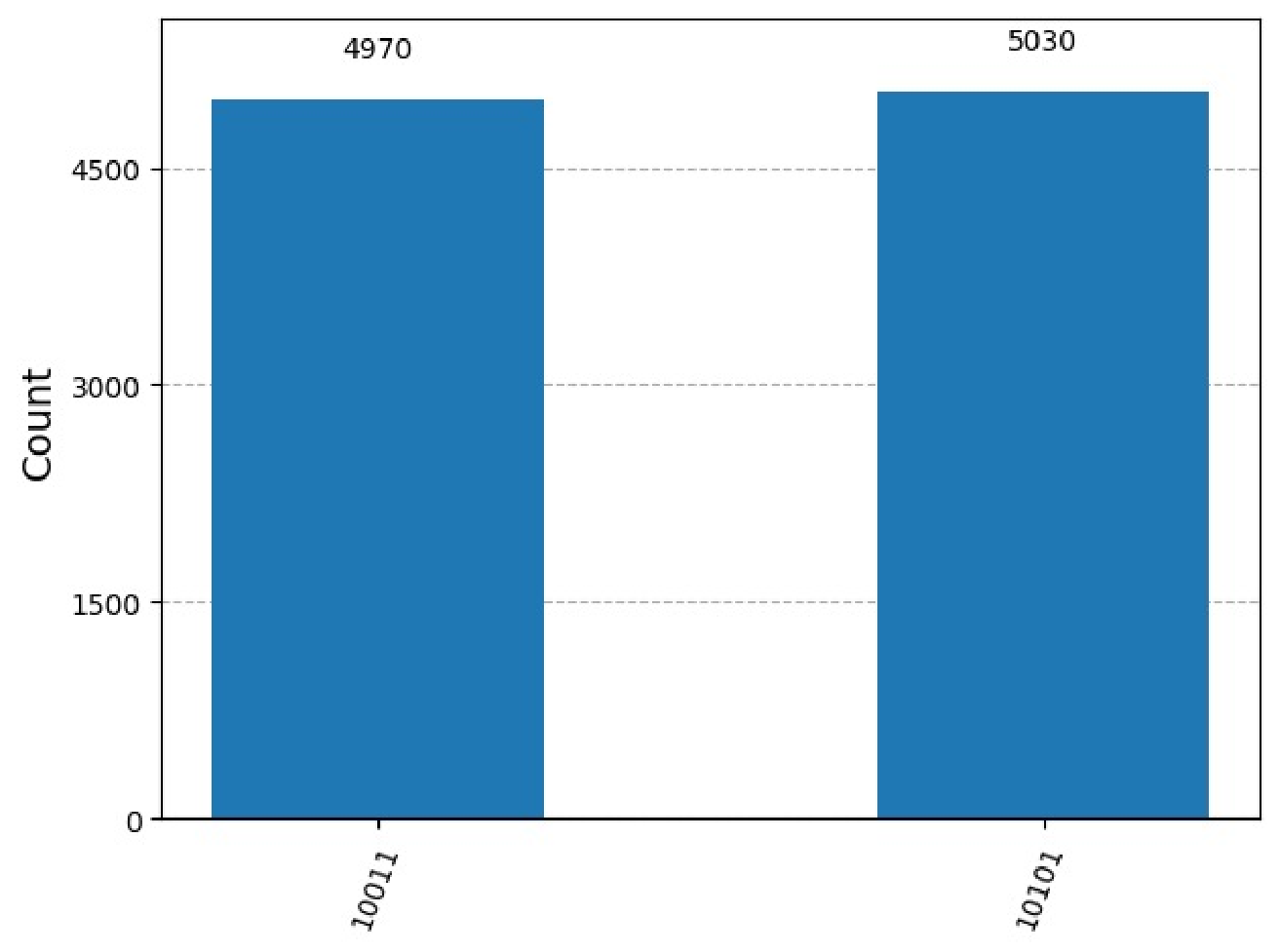}}
\end{center}
\caption{(a)  (a) Circuit employing Grover’s amplification to represent oracles $h^{(1)}(z)=\neg x_3 \wedge x_2   \wedge  x_1$, $h^{(2)}(y)=x_3 \wedge \neg x_2   \wedge  x_1$ for lower qubits, and the oracle for the upper qubits $ u(z)= x_5 \wedge \neg x_4$. When we measure, we are not guaranteed to obtain the solution $\xi=o(x)$ since $o(x)$ it was not checked and the possible combinations may be not present. (b) Due to the combinatorial problem that results in our case, yielding two solutions with equal probability, corresponding to the number of partial candidate solutions, since $(v=2)$ with  $\frac{1}{\sqrt{v}} \sum_{k=1}^v | u, h_k \rangle=\frac{1}{\sqrt{2}} \cdot  (| 10011 \rangle + | 10101 \rangle)$. }
\label{A_basic_0.eps}
\end{figure}

For  example in the preceding example we introduce with $v=2$ and $g=3$ 
\[ 
 h^{(1)}(y)=\neg x_3 \wedge x_2   \wedge  x_1,~~ h^{(2)}(z)=x_3 \wedge \neg x_2   \wedge  x_1
 \]
and the oracle for the  $(5-3)$ upper qubits
 \[
 u(z)= x_5 \wedge \neg x_4  
\]
as indicated by the circuit in Figure \ref{A_basic_0.eps} (a) with the  results of measurement shown in  Figure \ref{A_basic_0.eps} (b). 
The approach can be broadly characterized as
\begin{equation}
 \frac{1}{\sqrt{m-g}} \cdot  \sum_{z \in B^{m}}    (-1)^{u(z)} \cdot  | z \rangle  ,~~ \frac{1}{\sqrt{g}} \cdot  \sum_{y \in B^{g}}   (-1)^{h^{(k)}(y)} \cdot  | y \rangle  
\end{equation}
\begin{equation}
\left(  \frac{1}{\sqrt{m-g}} \cdot  \sum_{z \in B^{m}}  (-1)^{u(z)} \cdot  | z \rangle  \right)  \otimes  \left(  \frac{1}{\sqrt{g}} \cdot  \sum_{y \in B^{g}}   (-1)^{h^{(k)}(y)} \cdot  | y \rangle  \right)
\end{equation}
after applying Grover's algorithm to subspace   $\mathcal{L}$
\begin{equation}
\left(  \frac{1}{\sqrt{m-g}} \cdot  \sum_{z \in B^{m}}  (-1)^{u(z)} \cdot  | z \rangle  \right)   \otimes \left( \frac{1}{\sqrt{v}} \sum_{k=1}^v | h_k \rangle \right)
\end{equation}
and the subspace subspace   $\mathcal{U}$
\begin{equation}
| u \rangle  \otimes \left( \frac{1}{\sqrt{v}} \sum_{i=k}^v | h_k \rangle \right) =  \frac{1}{\sqrt{v}} \sum_{k=1}^v | u, h_k \rangle 
\end{equation}
since both subspaces  $\mathcal{L}$ and $\mathcal{U}$ are not entangled.

\subsection{Entangled Subspaces}

In original nested Grover's search  (see section 3.3),  after applying Grover's algorithm to the subspace $\mathcal{L}$ we get
\begin{equation}
  \left(  \frac{1}{\sqrt{m-g}} \cdot  \sum_{z \in B^{(m-g)}} | z \rangle  \right)   \otimes \left( \frac{1}{\sqrt{v}} \sum_{k=1}^v | h_k \rangle \right) 
\end{equation}
both subspaces  $\mathcal{L}$ and $\mathcal{U}$ are entangled because
\[ |x \rangle =  |z, y \rangle,   \]
\[ o(x) = o(z,y)= o(z,h_k)= u(z) \shortparallel h^{(k)}(y)= u(z) \shortparallel l(y),  \]
\begin{equation}
  \left( \frac{1}{\sqrt{m-g}} \cdot \frac{1}{ \sqrt{v}}  \cdot \sum_{z \in B^{(m-g)}}   \sum_{k=1}^v  (-1)^{o(z,h_k)} \cdot | z, h_k \rangle \right). 
\end{equation}
Applying Grover’s algorithm to the subspace $\mathcal{U}$ results in the creation of entanglement.
\begin{equation}
  \left(  \frac{1}{ \sqrt{m-g}} \cdot  \frac{1}{\sqrt { v}} \cdot  \sum_{z \neq u}   \sum_{h_k \neq l} | z, h_k \rangle \right)  + \frac{1}{\sqrt { v}}  |\xi \rangle.
\end{equation}
In our preceding example with $v=2$ (see see Figure \ref{A_basic_2.eps}) after applying Grover's algorithm to the subspace $\mathcal{L}$
\begin{equation}
 \frac{1}{2} \cdot \frac{1}{\sqrt{2}} \cdot    \sum_{z \in B^{2}}   \sum_{k=1}^2  (-1)^{o(z,h_2)} \cdot | z, h_k \rangle   
\end{equation}
\begin{equation}
  \left(  \frac{1}{\sqrt{8}} \cdot  \sum_{z \in B^{2}}    (-1)^{o(z,h_2)} \cdot | z, h_1 \rangle  \right)   + \left(  \frac{1}{\sqrt{8}} \cdot  \sum_{z \in B^{2}}  (-1)^{o(z,h_2)} \cdot  | z, h_2 \rangle  \right) 
\end{equation}
\begin{equation}
  \left( \frac{1}{\sqrt{8}} \cdot  \sum_{z \in B^{2}}    | z, h_1 \rangle  \right)   + \left(  \frac{1}{\sqrt{8}} \cdot  \sum_{z \in B^{2}}  (-1)^{o(z,h_2)} \cdot  | z, h_2 \rangle  \right) 
\end{equation}
After after applying Grover's algorithm to the subspace $\mathcal{U}$ we get
\begin{equation}
  \left(  \frac{1}{\sqrt{8}} \cdot  \sum_{z \in B^{2}}    | z, h_1 \rangle  \right)  +  \frac{1}{\sqrt{2}} \cdot  |\xi \rangle.
\end{equation}
With the state
\[  \ \frac{1}{\sqrt{8}} \cdot \left( | 00, h_1 \rangle  + | 01, h_1 \rangle + | 10, h_1 \rangle + | 11, h_1 \rangle \right) + \]
\[  \frac{1}{\sqrt{8}} \cdot \left( | 00, h_2 \rangle  - | 01, h_2 \rangle + | 10, h_2 \rangle + | 11, h_2 \rangle \right),  \]
after after applying Grover's algorithm to the subspace $\mathcal{U}$ we get
\[   \frac{1}{\sqrt{8}}\cdot \left( | 00, h_1 \rangle  + | 01, h_1 \rangle + | 10, h_1 \rangle + | 11, h_1 \rangle \right) + \frac{1}{\sqrt{2}} \cdot |\xi \rangle, \]
see Figure \ref{A_basic_2.eps} (b).
Grover’s Algorithm operates on the initial two qubits (upper qubits). It operates on two subspaces of $\mathcal{U}$, one for $h_1$ and the other for $h_2$. In the subspace defined by $h_1$, there is no marking. In the other subspace, there is a marking. As a result of Grover’s amplification, the first subspace exhibits a uniform distribution, while the second subspace contains a marking indicating the solution. 
Each of the two subspaces defined by $h_1$ and defined by $h_2$ has a probability of being measured of $0.5$. In the first subspace, the uniform distribution of each state represented by  is $0.5/4 =1/8 = 0.125$. In the second subspace, the solution is represented by $0.5$, as indicated in Figure \ref{A_basic_2.eps} (b). 
The number of partial candidate solutions, denoted as $v$, is inversely proportional to the probability of measuring the solution, which is represented as $1/v$. How can this problem be effectively addressed? Two potential solutions to this problem are the iterative approach and the idea of the disentanglement of the two subspaces subspace $\mathcal{L}$ and  $\mathcal{U}$ .

\section{The Iterative Approach}

\begin{figure}[htb]
\begin{center}
\leavevmode
\epsfysize10cm
\epsffile{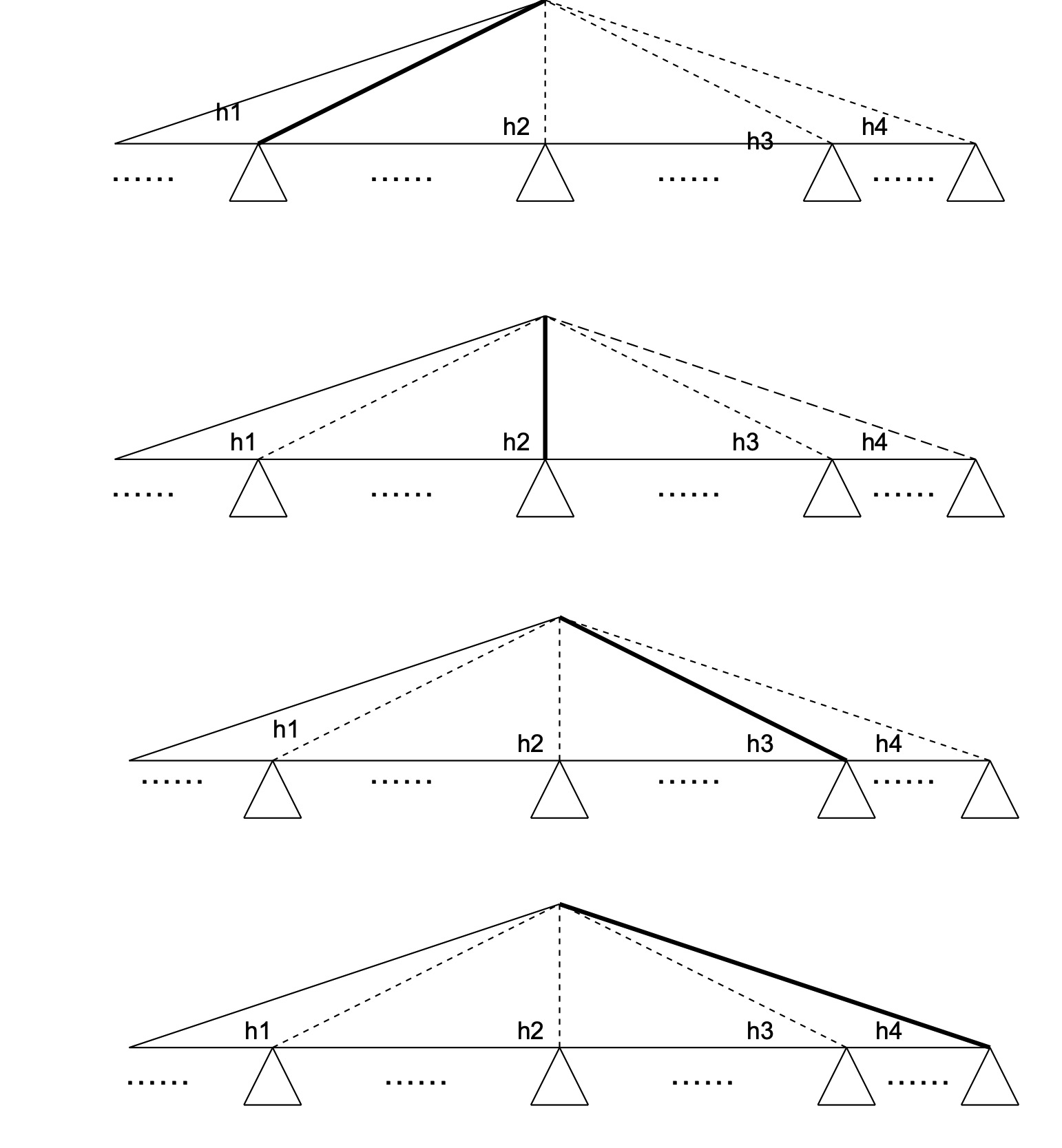}
\end{center}
\caption{ Example for $v=4$. Determine the solution $h_1$  by Grover algorithm on the subspace $\mathcal{L}$. Verify if the solution indicated by the oracle  $ o_{\xi}(x)$ exists. Repeat the procedure for $h_2, h_3, h_4.$
} \label{iterative.eps}
\end{figure}

\begin{figure}
\leavevmode
\parbox[b]{13cm}{ (a) \epsfxsize13cm\epsffile{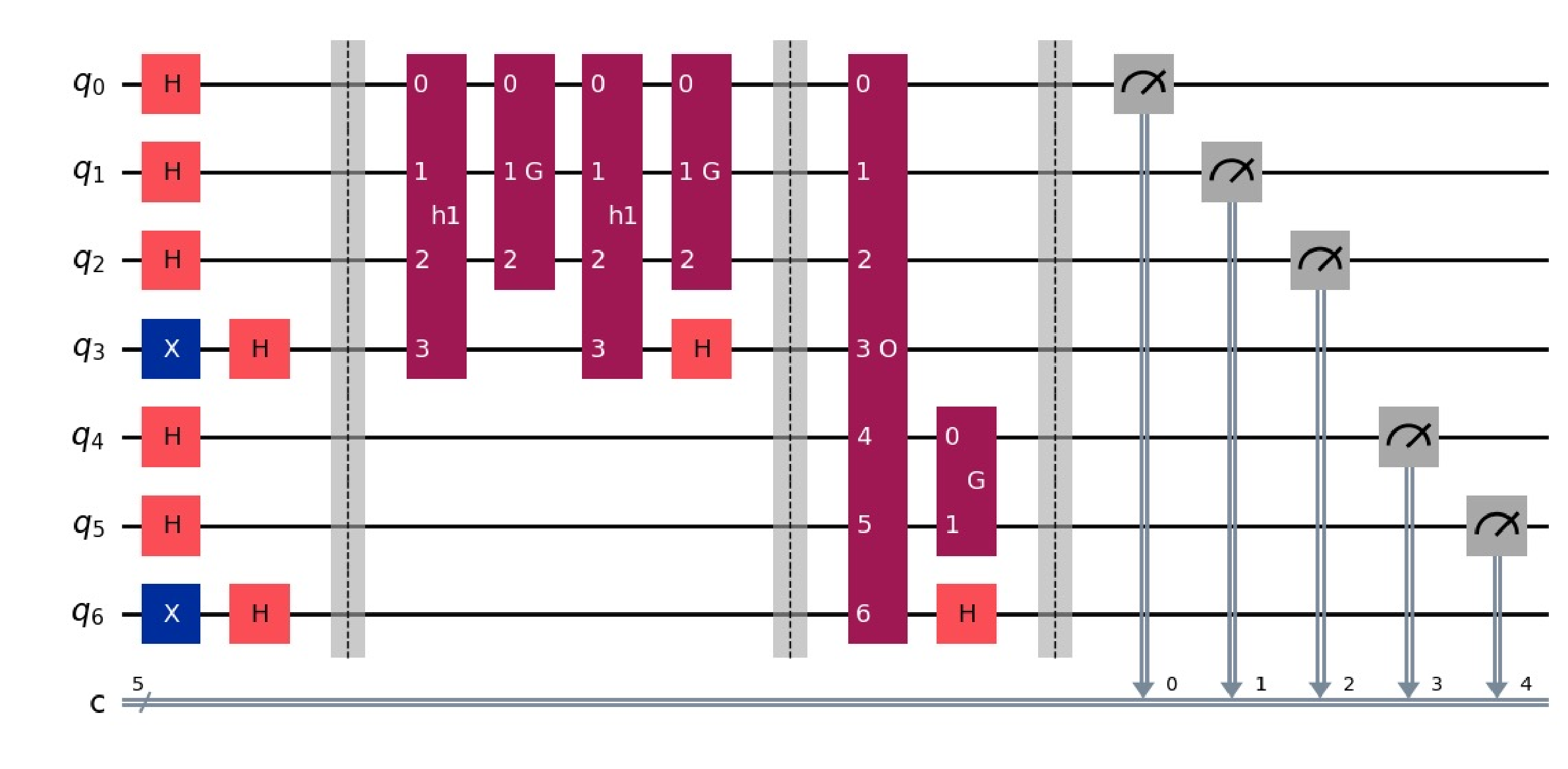}}
\parbox[b]{7cm}{ (b) \epsfxsize6cm\epsffile{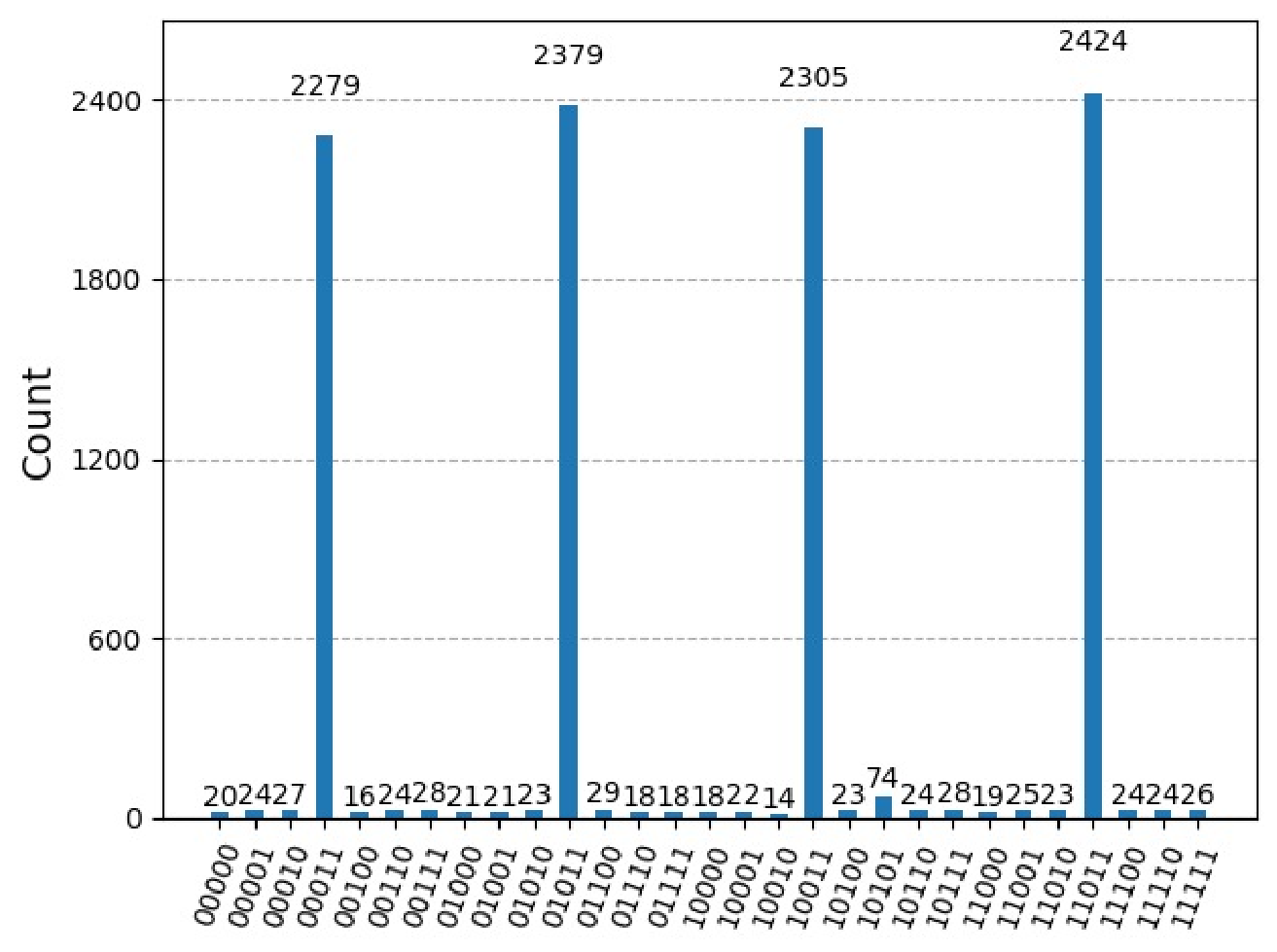}}
\parbox[b]{7cm}{ (c) \epsfxsize6cm\epsffile{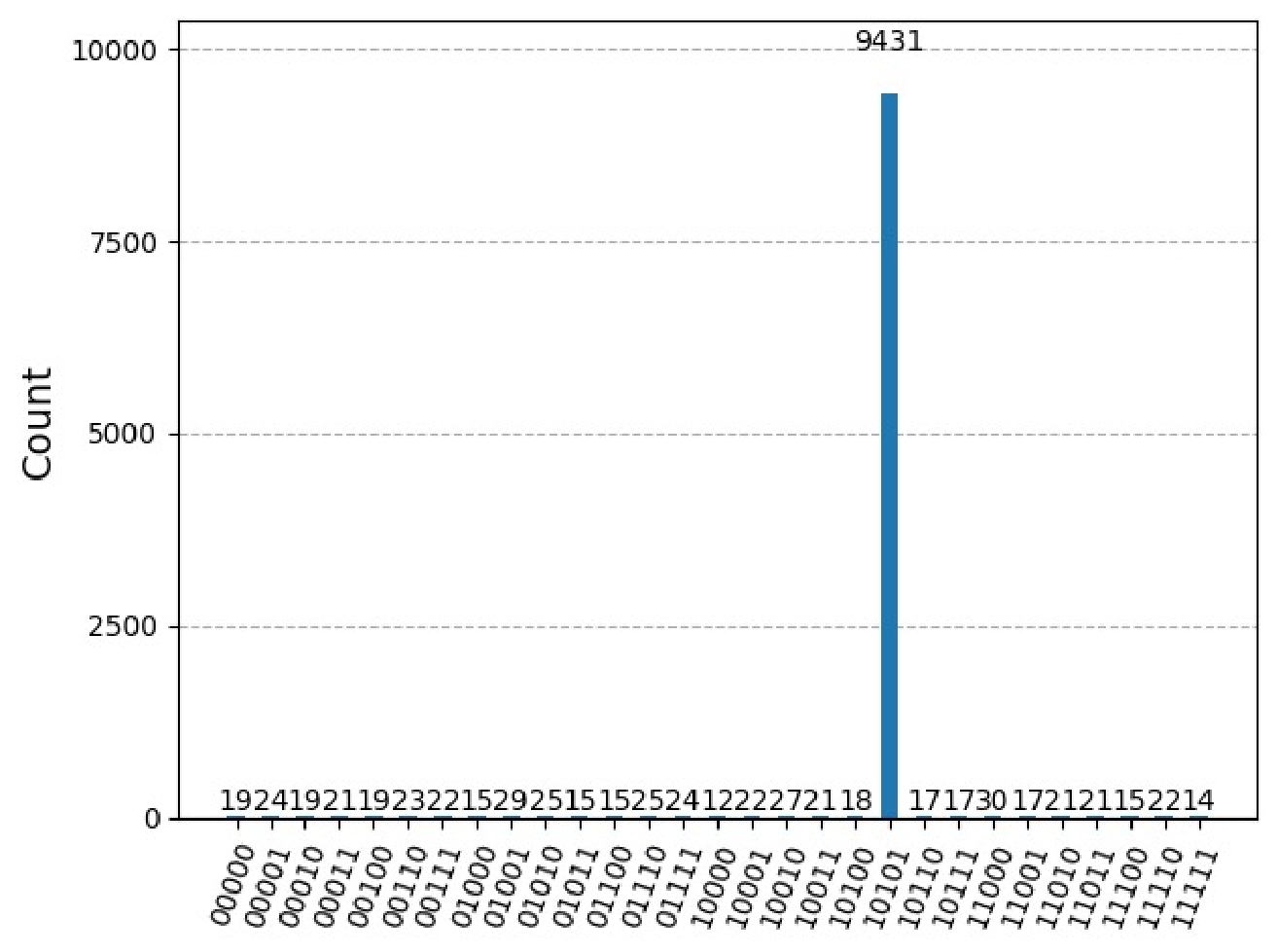}}
\caption{(a) Circuit using Grover's amplification representing oracle $h^{(1)}(y)=\neg x_3 \wedge x_2   \wedge  x_1$ for lower qubits and the global oracle $ o(x)= x_5 \wedge \neg x_4  \wedge    x_3 \wedge \neg x_2   \wedge  x_1$. (b) Since the solution does not exist, when we measuring we get the combination of $h_1$ with all four possible combination of the two upper qubits $x_5$ an $x_6$ with the probability $1/8$. (c) The solution for a circuit  $\xi = u \shortparallel l$ with $l=h_2$.   It is the same circuit as shown in (a), only  $h^{(1)}(y)$ is replaced with $h^{(2)}(z)=x_3 \wedge \neg x_2   \wedge  x_1$.   } 
\label{A_basic_4.eps}
\end{figure}

With the  set $\aleph$  of $v$  partial candidate solutions 
\[
  h^{(k)}(y)=h_k~~with~ k \in \{ 1, 2,\cdots , v   \}
\]
and  the global oracle $ o_{\xi}(x)$, we built a circuit with one possible could-be solution of the set $\aleph$ and the global oracle.
We determine in the first step  one solution of  one $ h^{(k)}(y)$  by Grover algorithm on the subspace $\mathcal{L}$, in the second step we  the global oracle  $ o_{\xi}(x)$  acts on $m$ qubits. Subsequently, we apply Grover’s amplification to the subspace $\mathcal{U}$, which is defined by the upper $m-g$ qubits. 
Figure \ref{iterative.eps} indicates an example representing the decomposed search tree for $v=4$.
If a solution exists, the subspaces $\mathcal{L}$ and $\mathcal{U}$ are not entangled with $\mathcal{H}=\mathcal{U} \otimes \mathcal{L}$.
In this case, by measuring $\mathcal{H}$, we obtain the solution.
This is because the number of partial candidate solutions represented in our circuit is “one,” $v^*=1,$ and the probability of measuring a solution if it exists is $1/v^*=1.$
We measure the state and test whether a solution is present. If it is, we terminate the process; otherwise, we select another potential solution from the set $\aleph.$ 
For  the simple preceding SAT example with $v=2$ and $g=3$ and $m=5$ we get two circuits, see Figure \ref{A_basic_4.eps}.

 \subsection{Algorithm}
 
Given set $\aleph$  of $v$  partial candidate solutions 
\[
  h^{(k)}(y)=h_k~~with~ k \in \{ 1, 2,\cdots , v   \}
\]
and  the global oracle $ o_{\xi}(x)$.

\begin{enumerate}
\item For $k=1$ to $v$ 
\item ~~~For could-be solution $h^{(k)}(y)=h_k$ built a circuit;
\item  ~~~Determine the solution $h_k$  by Grover algorithm on the subspace $\mathcal{L}$;
\item  ~~~Verify if the solution indicated by the oracle  $ o_{\xi}(x)$ exists;
\item  ~~~If solution exists success, exit the loop otherwise continue the loop;
\end{enumerate}
Since both subspaces $\mathcal{L}$ and $\mathcal{U}$ are not entangled, if a solution exists, it is readily apparent.
We measure the state and verify whether it is a valid solution. If it is a solution, we conclude the process. Otherwise, we select another  partial candidate solution from the set $\aleph$.

\subsection{Optimal Dimension}

How to choose an optimal dimension of the two subspaces $\mathcal{U}$ and $\mathcal{L}$?  The optimal dimension of the two subspaces results in the minimal costs represented  by the minimum of the function
\begin{equation}
\sqrt{2^{m\cdot(1-a)}} + \sqrt{2^{{m \cdot a}}}~~with~ a \in [0,1]  
\end{equation}
which is equivalent to 
\[ e^{m\cdot(1-a)}+e^{m\cdot a}~~with~ a \in [0,1]  \]
\begin{equation}
\frac{\partial}{\partial a} \cdot \left( e^{m\cdot(1-a)}+e^{m\cdot a} \right) =  - e^{m\cdot(1-a)} \cdot m+e^{m\cdot a} \cdot m ~~with~ a \in [0,1]  
\end{equation}
with the solution
\[a=\frac{1}{2}. \]

\subsection{Costs}

Within the subspace $\mathcal{U}$, the costs associated with Grover’s algorithm are $\sqrt{2^{m/2}}$, while within the subspace $\mathcal{L}$, the corresponding cost remains $\sqrt{2^{m/2}}$. We need to verify whether the proposed solution. Subsequently, we must repeat the procedure in the worst case $v$ times leading to the cost
\begin{equation}
 v \cdot \left(\sqrt{2^{m/2}} + \sqrt{2^{{m/2}}} + 1 \right) =   v \cdot \left(2 \cdot  2^{{m/4}} +1  \right)
 \end{equation} 
with the constraint for $v$ that the cost are  below $2^{m/2}$ with
\begin{equation}
 v < \frac{n^ {1/2}}   {2  \cdot  n^ {1/4}+ 1}  \approx n^ {1/4}=2^{m/4}.
 \label{v:const}
 \end{equation} 
 since $n=2^m$.
 We observe that maximal possible value for  $v$ is approximately equal to the number of nodes located at one-fourth of the depth of the search tree. 
We can express the cost of iterative search it in $O$ notation given $v \approx m$  as 
\begin{equation}
 O(v \cdot  n^{1/4})=O(\log n \cdot  n^{1/4}).
  \end{equation}

\section{Disentanglement of $\mathcal{L}$ and $\mathcal{U}$}

Can we disentangle the spaces $\mathcal{L}$ and $\mathcal{U}$ with  $\mathcal{H}=\mathcal{U}   \otimes \mathcal{L}$  by knowing the function $u(z)$?  We test the global oracle  $o_{\xi}(x)$ over the whole space of dimension $m$ that defines the Hilbert space $n=2^m$.
The test should not entangle both sub spaces  $\mathcal{L}$ and  $\mathcal{U}$. We really on the assumption that $\xi $ exists and can be decomposed as
\[\xi = u \shortparallel  l. \]
 Knowing  $o_{\xi}(x)$, assuming the  $u(z)$ is correct and with set $\aleph$ the task is to verify if 
 \[ o(x) = o(z,h_k)=  u(z) \shortparallel l(h_k)  \]
  is true.
 If it is true, then we can decompose after Grover's algorithm on the subspace $\mathcal{L}$
 \begin{equation}
  \frac{1}{\sqrt{2^m}} \frac{1}{\sqrt{v}} \cdot  \sum_{z \in B^{m}}   \sum_{k=1}^v (-1)^{o(z,h_k)} \cdot | z, h_k \rangle  =
\end{equation}
 \begin{equation}
 \frac{1}{\sqrt{2^m}} \frac{1}{\sqrt{v}} \cdot   \sum_{z \in B^{m}}   \sum_{k=1}^v (-1)^{o(z,h_k)} \cdot | z, h_k \rangle =
\end{equation}
 \begin{equation}
 \frac{1}{\sqrt{2^m}} \frac{1}{\sqrt{v}}  \cdot    \sum_{z \in B^{m}}  \sum_{ h_k \neq l} (-1)^{o(z,h_k)} \cdot  | z, h_k \rangle  +  \frac{1}{\sqrt{2^m}} \frac{1}{\sqrt{v}}  \cdot  (-1)^{o(z,h_k)} \cdot |\xi \rangle =
 \end{equation}
 \begin{equation}
 \frac{1}{\sqrt{2^m}} \frac{1}{\sqrt{v}} \cdot   \sum_{z \in B^{m}}  \sum_{ h_k \neq l}   | z, h_k \rangle  -  \frac{1}{\sqrt{2^m}} \frac{1}{\sqrt{v}} \cdot  |\xi \rangle 
\end{equation}
After Grover's search on the subspace $\mathcal{U}$ if true
\begin{equation}
 \frac{1}{\sqrt{v}} \cdot \left(| u \rangle  \otimes  \sum_{h_k \neq l}| h_k \rangle \right) +  \frac{1}{\sqrt{v}} \cdot |\xi \rangle  = 
 \end{equation}
 \[
 | u \rangle  \otimes \left( \frac{1}{\sqrt{v}} \sum_{i=k}^v | h_k \rangle \right)
\]
If not true for $v > 2$
 \[
\frac{1}{\sqrt{v-2}} \cdot    | u \rangle \otimes \frac{1} {\sqrt{(v-2) \cdot 2^m}} \left( \sum_{z \in B^{m}}   | z \rangle \right)   \otimes \left( \frac{1}{\sqrt{v}} \sum_{i=k}^v | h_k \rangle \right)
\]
and for  $v = 2$
 \[
\frac{1} {\sqrt{ 2^m}}  \left( \sum_{z \in B^{m}}   | z \rangle \right)   \otimes \left( \frac{1}{\sqrt{v}} \sum_{i=k}^v | h_k \rangle \right).
\]
In the case $| u \rangle$ does not exist for any $z$
 \begin{equation}
  \left(\frac{1}{\sqrt{2^m}}  \frac{1}{\sqrt{v}}  \cdot  \sum_{z \in B^{m}}   \sum_{k=1}^v  | z, h_k \rangle \right) = \left( \frac{1}{\sqrt{2^m}}  \sum_{z \in B^{m}}  | z  \rangle \right) \otimes \left( \frac{1}{\sqrt{v}}    \sum_{k=1}^v  | h_k \rangle \right).
\end{equation}
Since we do not know which $h_k=l$ we have to test all combinations, but we can reused the set $\aleph$ once computed.
\[
  \left( \frac{1}{\sqrt{2^m}} \frac{1}{\sqrt{v}} \cdot  \sum_{z \in B^{m}}  \sum_{ h_k \neq h_1}   (-1)^{o(z,h_k)}  \cdot  | z, h_k \rangle  +     \frac{1}{\sqrt{2^m}} \frac{1}{\sqrt{v}} \cdot  (-1)^{o(z,h_k)}  \cdot |\xi \rangle \right)
\]
\[
\otimes  \left(  \frac{1}{\sqrt{2^m}} \frac{1}{\sqrt{v}} \cdot    \sum_{z \in B^{m}}  \sum_{ h_k \neq h_2}   (-1)^{o(z,h_k)}  \cdot  | z, h_k \rangle  +     \frac{1}{\sqrt{2^m}} \frac{1}{\sqrt{v}} \cdot   (-1)^{o(z,h_k)}  \cdot |\xi \rangle  \right)
\]
\[
\cdots \cdots \cdots \cdots
\]
 \begin{equation}
\otimes  \left( \frac{1}{\sqrt{2^m}} \frac{1}{\sqrt{v}}  \cdot    \sum_{z \in B^{m}}  \sum_{ h_k \neq h_v}  (-1)^{o(z,h_k)}  \cdot  | z, h_k \rangle  +  \frac{1}{\sqrt{2^m}} \frac{1}{\sqrt{v}}  \cdot     (-1)^{o(z,h_k)}  \cdot |\xi \rangle \right)=
\end{equation}
\[
  \left( \frac{1}{\sqrt{2^m}} \frac{1}{\sqrt{v}} \cdot  \sum_{z \in B^{m}}  \sum_{ h_k \neq h_1}   | z, h_k \rangle  -     \frac{1}{\sqrt{2^m}} \frac{1}{\sqrt{v}}  \cdot |\xi \rangle \right)
\]
\[
\otimes  \left(  \frac{1}{\sqrt{2^m}} \frac{1}{\sqrt{v}} \cdot    \sum_{z \in B^{m}}  \sum_{ h_k \neq h_2}   | z, h_k \rangle  -    \frac{1}{\sqrt{2^m}} \frac{1}{\sqrt{v}} \cdot |\xi \rangle  \right)
\]
\[
\cdots \cdots \cdots \cdots
\]
 \begin{equation}
\otimes  \left( \frac{1}{\sqrt{2^m}} \frac{1}{\sqrt{v}}  \cdot    \sum_{z \in B^{m}}  \sum_{ h_k \neq h_v}   | z, h_k \rangle  -  \frac{1}{\sqrt{2^m}} \frac{1}{\sqrt{v}}  \cdot      |\xi \rangle \right).
\end{equation}
After  after Grover's algorithm on the subspace $\mathcal{U}$
\[
\left( \left(  \frac{1}{\sqrt{v-2}} \cdot    | u \rangle + \frac{1} {\sqrt{(v-2) \cdot 2^m}} \sum_{z \in B^{m}}   | z \rangle \right)   \otimes \left( \frac{1}{\sqrt{v}} \sum_{i=k}^v | h_k \rangle \right) \right)
\]
 \[
\otimes \left(  | u \rangle  \otimes \left( \frac{1}{\sqrt{v}} \sum_{i=k}^v | h_k \rangle \right) \right)
\]
\[
\cdots \cdots \cdots \cdots
\]
\[
\otimes \left( \left(  \frac{1}{\sqrt{v-2}} \cdot    | u \rangle + \frac{1} {\sqrt{(v-2) \cdot 2^m}} \sum_{z \in B^{m}}   | z \rangle \right)   \otimes \left( \frac{1}{\sqrt{v}} \sum_{i=k}^v | h_k \rangle \right)  \right)
\]
and if $u$ not present we get a general superposition. We arrive at the decomposition
\begin{equation}
\mathcal{H}=\mathcal{U}_{ h_k \neq h_1}    \otimes \mathcal{U}_{ h_k \neq h_2}   \otimes \cdots  \otimes \mathcal{U}_{ h_k \neq h_v}   \otimes \mathcal{L}
\end{equation}
We have determined the correct $h^{k}(y)$ by measuring $| u \rangle$ with a probability of approximately 1 in $\mathcal{U}$. Subsequently, we can estimate the path descriptor $h_k=l$ using Grover’s algorithm on the subspace $\mathcal{L}$. The probability measurement of $\frac{1}{v-2}$ indicates that $h^{k}(y)$ does not equal $l$ and does not constitute a valid solution. However, for $v > 4$, it necessitates a substantial number of measurements to ascertain that the probability deviates from $1$.

 \subsection{Example with $v=2$}

Our previous example with $v=2$
\[
  \left( \frac{1}{\sqrt{2^3}}  \frac{1}{\sqrt{2}} \cdot     \sum_{z \in B^{3}}   (-1)^{o(z,h_k)}  \cdot  | z, h_1 \rangle  +   \frac{1}{\sqrt{2^3}}  \frac{1}{\sqrt{2}} \cdot  (-1)^{o(z,h_k)}  \cdot |\xi \rangle \right) 
\]
\[
\otimes  \left(\frac{1}{\sqrt{2^3}}  \frac{1}{\sqrt{2}} \cdot    \sum_{z \in B^{3}}   (-1)^{o(z,h_k)}  \cdot  | z, h_2 \rangle  +   \frac{1}{\sqrt{2^3}}  \frac{1}{\sqrt{2}} \cdot  (-1)^{o(z,h_k)}  \cdot |\xi \rangle \right)=
\]
\[
  \left( \frac{1}{\sqrt{2^3}}  \frac{1}{\sqrt{2}} \cdot     \sum_{z \in B^{3}}    | z, h_1 \rangle  -   \frac{1}{\sqrt{2^3}}  \frac{1}{\sqrt{2}} \cdot   |\xi \rangle \right) 
\]
\[
\otimes  \left(\frac{1}{\sqrt{2^3}}  \frac{1}{\sqrt{2}} \cdot    \sum_{z \in B^{3}}   | z, h_2 \rangle  -  \frac{1}{\sqrt{2^3}}  \frac{1}{\sqrt{2}} \cdot  |\xi \rangle \right).
\]
After Grover's algorithm  on the subspace  $\mathcal{U}$ we get the decomposition
\begin{equation}
\mathcal{H}=\mathcal{U}_{ h_k \neq h_1}    \otimes \mathcal{U}_{ h_k \neq h_2}    \otimes \mathcal{L}=\mathcal{U}_{  h_2}    \otimes \mathcal{U}_{  h_1}    \otimes \mathcal{L},
\end{equation}
with
\[
  \left( \frac{1} { 2^3}  \sum_{z \in B^{m}}   | z \rangle  \right)  \otimes  \left( \frac{1}{2} \cdot \left(  |\xi \rangle  + | u \rangle | h_2 \rangle \right)   \right)   \otimes \left(\frac{1}{2} \cdot \left(  |h_1 \rangle  + | h_2 \rangle\right)  \right)=
\]
 \[
   \left( \frac{1} { 2^3}  \sum_{z \in B^{m}}   | z \rangle  \right)   \otimes \left( \frac{1}{2} \cdot \left(   | u \rangle | h_1 \rangle  + | u \rangle | h_2 \rangle \right)   \right) \otimes \left(\frac{1}{2} \cdot \left(  |h_1 \rangle  + | h_2 \rangle\right)  \right)=
\]
 \[
  \left( \frac{1} { 2^3}\sum_{z \in B^{m}}   | z \rangle   \right)   \otimes |u \rangle \otimes \frac{1}{2} \cdot \left(  |h_1 \rangle  + | h_2 \rangle\right). 
\]
Because of the equality $|\xi \rangle  = | u \rangle | h_1 \rangle$ we know that $h^{1}(y)$  is correct  and can estimate  the path descriptor $h_1=l$ by Grover's algorithm on the subspace $\mathcal{L}$.

\subsection{Costs}

Within the subspace $\mathcal{L}$, the cost associated with Grover’s algorithm is $\sqrt{(2^{m/2}/v)}$. For $v$ combinations in the subspace $\mathcal{U}$, the corresponding cost is $v \cdot \sqrt{2^{{m/2}}}$. Additionally, given $h_k$, we need to determine the path descriptor $h_k = l$ by Grover’s algorithm in the subspace $\mathcal{L}$. The complete costs are
\begin{equation}
\sqrt{ \frac{2^{m/2}}{v}} +  \sqrt{2^{m/2}} +   v \cdot \sqrt{2^{{m/2}}} =  \sqrt{2^{{m/2}}} \cdot \left( \frac{1}{ \sqrt{v}} + 1 +v  \right)
 \end{equation} 
 We can express the cost of iterative search it in $O$ notation given $v \approx m$ as before with
\[ O(v \cdot  n^{1/4})=O(\log n \cdot  n^{1/4}).\]
For $1 < v$, the actual costs are less than the cost of the iterative algorithm with
\begin{equation}
\sqrt{ \frac{2^{m/2}}{v}} +  \sqrt{2^{m/2}} +   v \cdot \sqrt{2^{{m/2}}}  <      v \cdot \left(\sqrt{2^{m/2}} + \sqrt{2^{{m/2}}} + 1 \right) 
 \end{equation} 
\begin{equation}
  \sqrt{2^{m/2}}  \cdot \left( \frac{1}{\sqrt{v} }   + 1 \right)       <        v \cdot  \sqrt{2^{m/2}}    <    v \cdot \left( \sqrt{2^{m/2}} + 1 \right)
 \end{equation}
 \begin{equation}
  \frac{1}{\sqrt{v} }   + 1 < v,
  \end{equation}
they are cheaper in relation $times$.
\begin{equation}
times=\frac{   v \cdot \left(2 \cdot \sqrt{2^{m/2}}  + 1 \right) } {  \sqrt{2^{{m/2}}} \cdot \left( \frac{1}{ \sqrt{v}} + 1 +v  \right) },
 \end{equation}
For $v > 4$, a substantial number of measurements is necessary. A practical approach would be to select $v^* = 4$, resulting in approximately $1.45$ faster costs
\[
times=\frac{  8 \cdot \sqrt{2^{m/2}}  + 4  } {  \sqrt{2^{{m/2}}} \cdot 5.5  }  \approx 1.45
\]
compared to the iterative approach with $v/4=v/v^*$ possible iterations. 

\subsubsection{Implementation}

For the $\aleph$  of $v$  partial candidate solution functions that acts on $g$ lower qubits  on the subspace  $\mathcal{L}$
\[
  h^{(k)}(y)~~with~ k \in \{ 1, 2,\cdots , v   \}
\]
we introduce a $flag_k$ variable
\begin{equation}
  h^{(k)}(y)= \left\{
  \begin{array}{l} 
1=flag_k~~~ if~~~y=h_k \\
 0=flag_k~~~else
 \end{array}  \right. 
 \end{equation}
and extend the upper oracle in relation to the  $flag_k$  variable
\begin{equation}
  \tilde{u}^{(k)}(z)= \left\{
  \begin{array}{l} 
1~~~ if~~~(z=u) \wedge (flag_k=1)  \\
 0~~~else
 \end{array}  \right.
 \end{equation}
 For example, for the case $l=h_s$, $\tilde{u}^{(s)}(z)=o_{\xi}(x)$. Do determine a possible solution we have to measure $v$ subspaces $\mathcal{U}$.
  
\begin{figure}
\leavevmode
\parbox[b]{13cm}{ (a) \epsfxsize14cm\epsffile{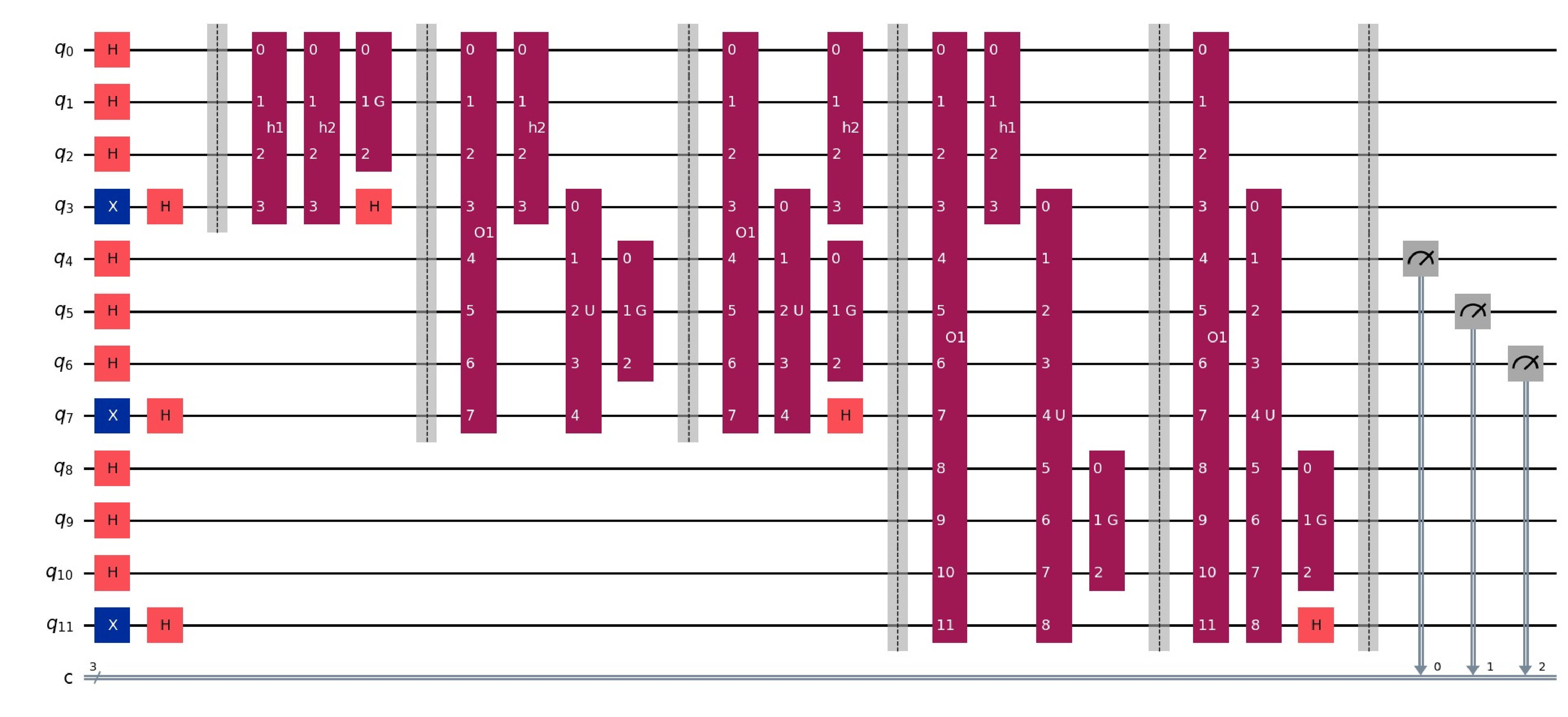}}
\parbox[b]{8cm}{ (b) \epsfxsize6cm\epsffile{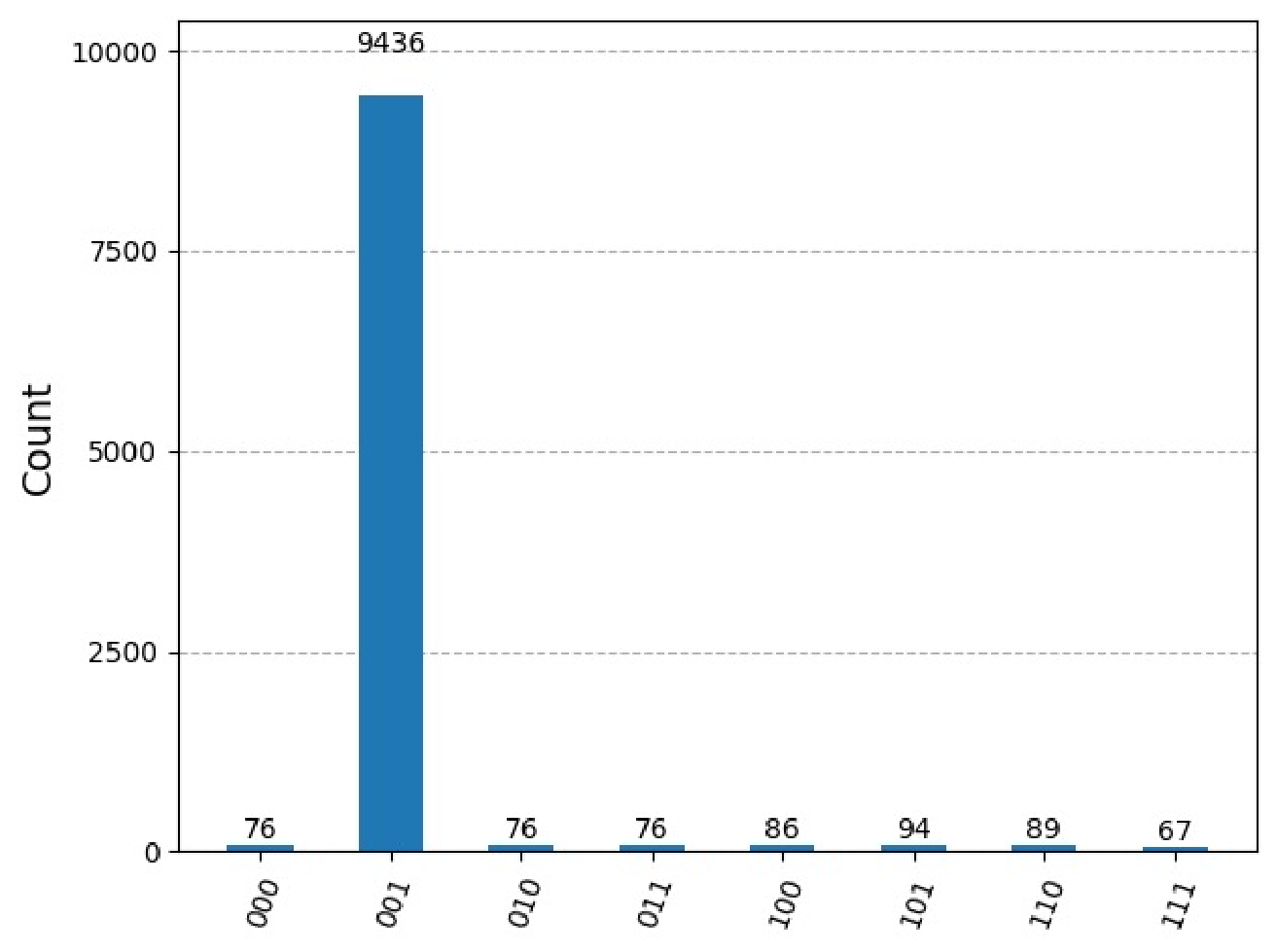}}
\parbox[b]{8cm}{ (c) \epsfxsize6cm\epsffile{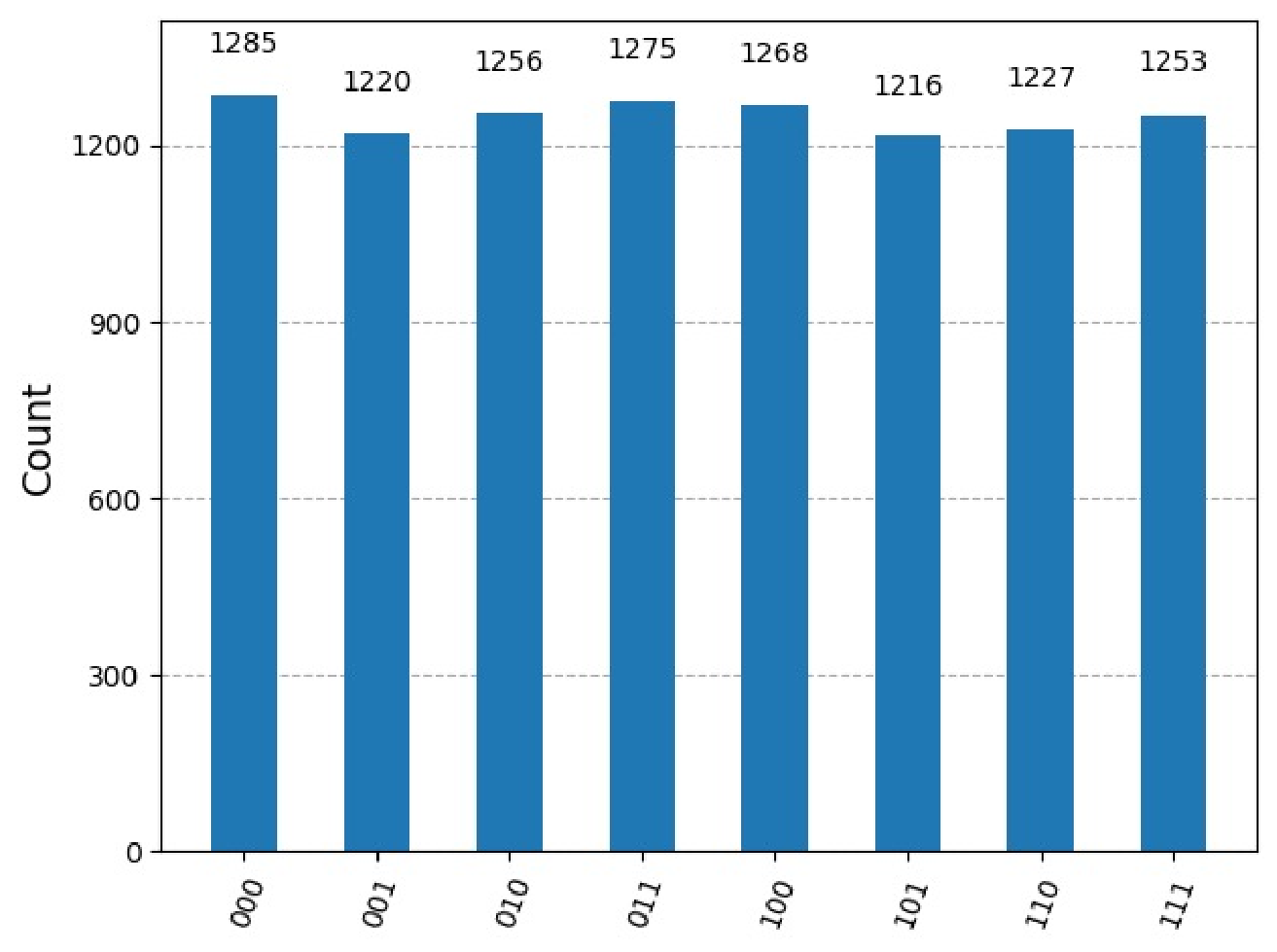}}
\caption{(a) Circuit employing Grover’s amplification to represent oracles $h^{(1)}(z)=\neg x_3 \wedge x_2   \wedge  x_1$, $h^{(2)}(z)=x_3 \wedge \neg x_2   \wedge  x_1$ representing the decomposition  $\mathcal{U}_{  h_2}    \otimes \mathcal{U}_{  h_1}    \otimes \mathcal{L}$.
The subspace $\mathcal{L}$ is represented by the qubits $0-2$,  subspace $\mathcal{U}_{  h_1}$ is represented by the qubits $4-6$ and  subspace $\mathcal{U}_{  h_2} $ is represented by the qubits $8-10$.
 (b) We measure  the subspace $\mathcal{U}_{  h_1}$  representing the solution. (c) We measure  the subspace $\mathcal{U}_{  h_2}$  representing a non  solution. } 
\label{A_basic_10_1st.eps}
\end{figure}

 For  example in the preceding SAT example  with $v=2$ and $g=3$ we introduce   the  $flag_k$  variable
\[ 
 h^{(1)}(y)=\neg x_3 \wedge x_2   \wedge  x_1,~~ h^{(2)}(z)=x_3 \wedge \neg x_2   \wedge  x_1
 \]
  \[
 u(z)^{(k)}= (\neg x_6 \wedge  \neg x_5    \wedge x_4 ) \wedge flag_k
\]
\[
 o(x)= \neg x_6 \wedge  \neg x_5    \wedge x_4 \wedge  \neg x_3  \wedge x_2 \wedge  x_1
 \]
where  $o_{\xi}(x)$ corresponds to $ u(y)^1$. The decomposition
\[
\mathcal{H}=\mathcal{U}_{  h_2}    \otimes \mathcal{U}_{  h_1}    \otimes \mathcal{L},
\]
is represented by the quantum circuit  in Figure  \ref{A_basic_10_1st.eps} (a). 

\section{Permutation in the subspace $\mathcal{L}$ }

\begin{figure}
\leavevmode
\parbox[b]{7cm}{ (a) \epsfxsize7cm\epsffile{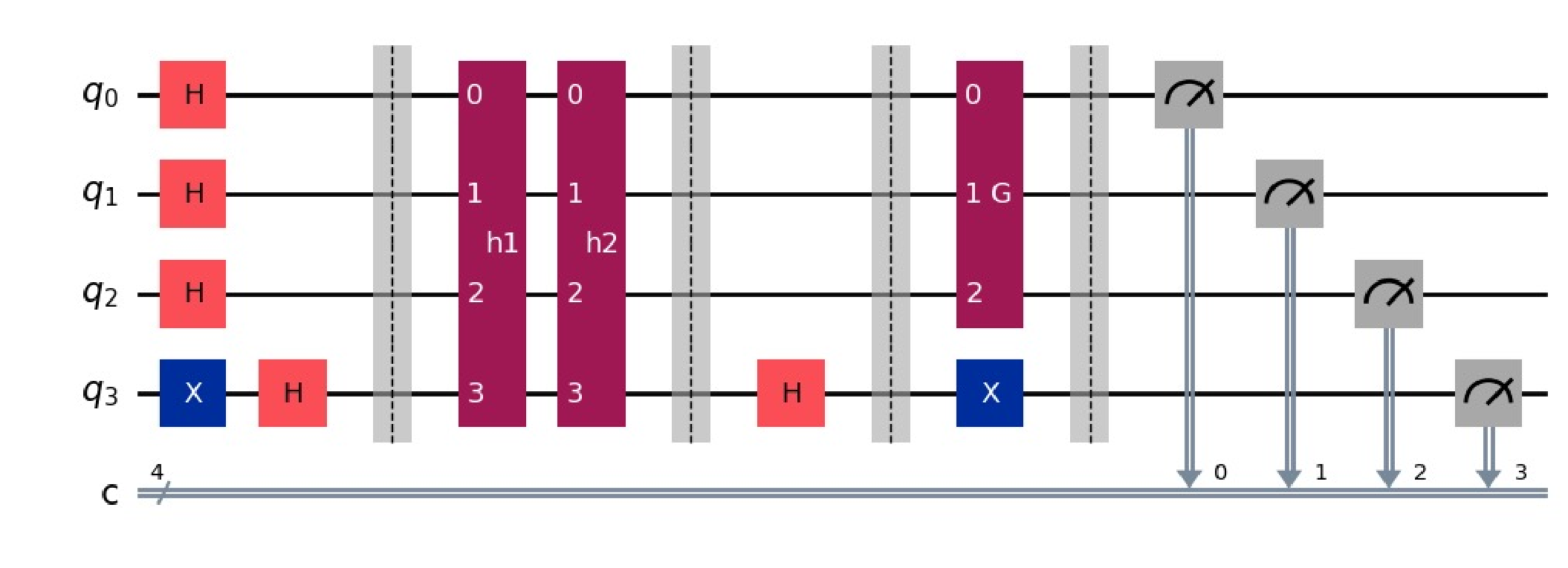}}
\parbox[b]{7.5cm}{ (b) \epsfxsize7cm\epsffile{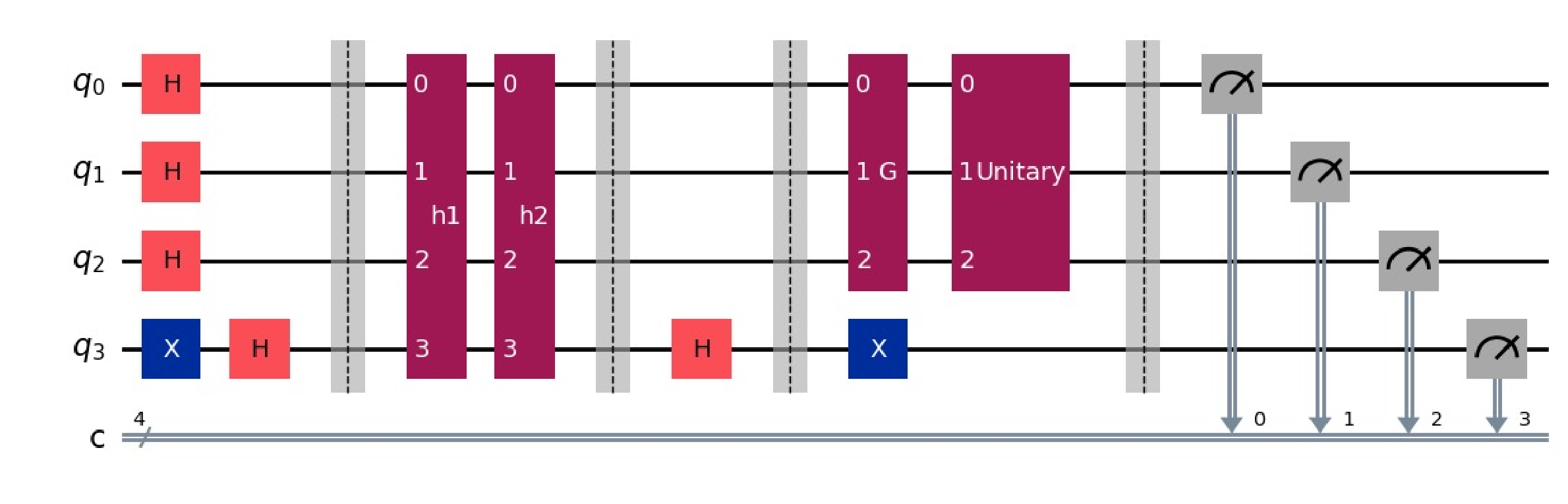}}
\parbox[b]{7cm}{ (c) \epsfxsize6cm\epsffile{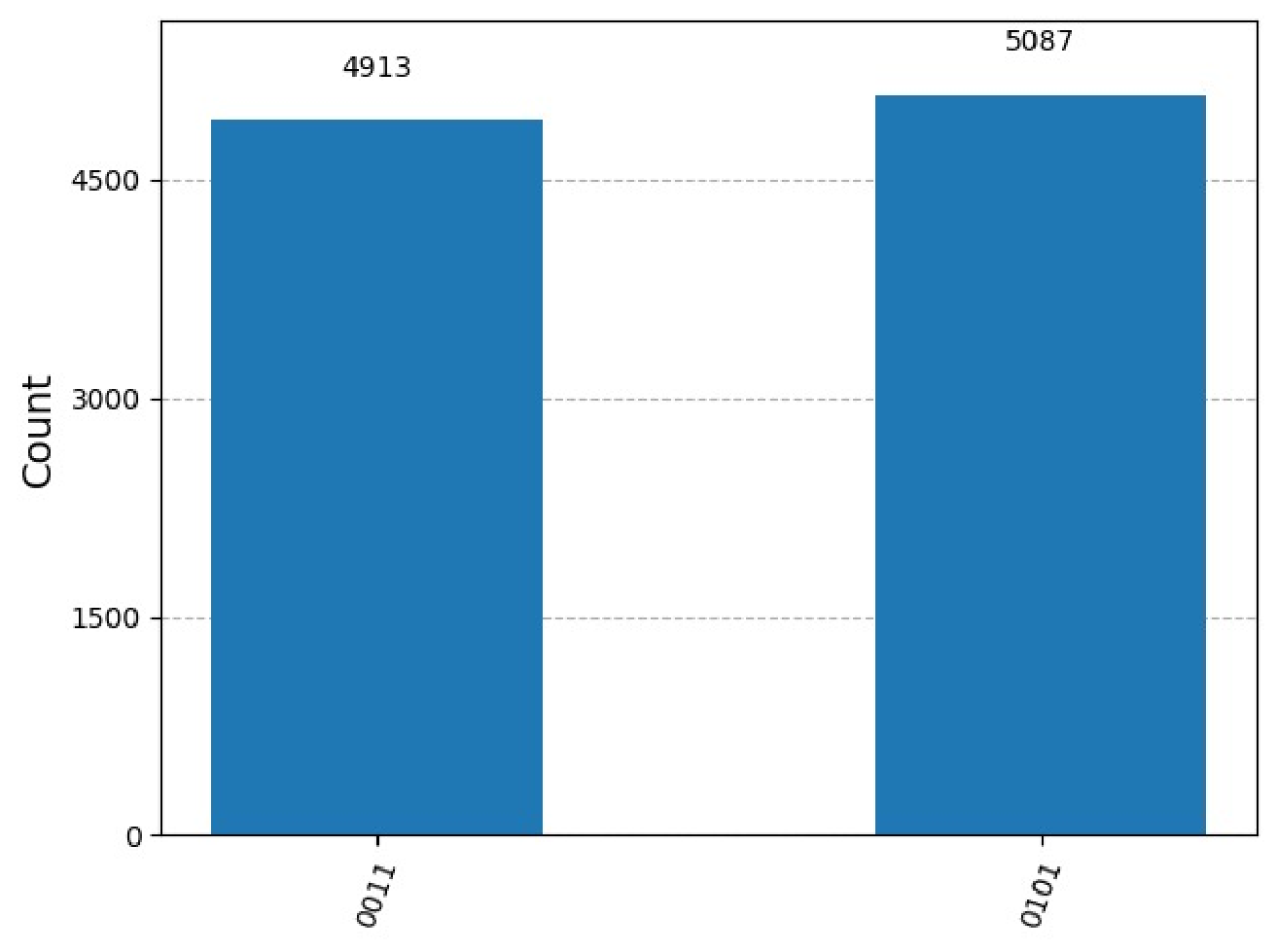}}
\parbox[b]{7cm}{ (d) \epsfxsize6cm\epsffile{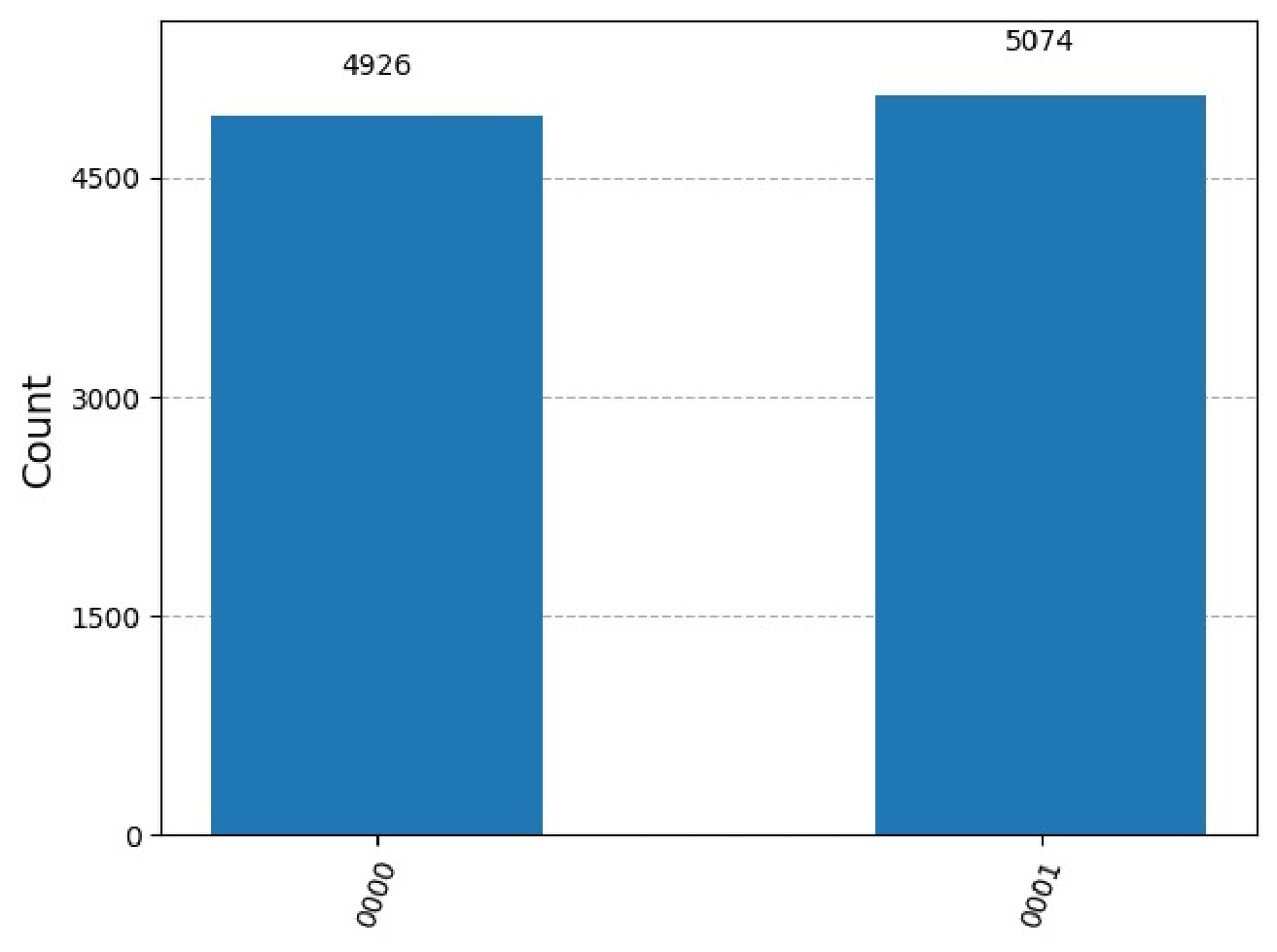}}
\caption{(a) After applying Grover's algorithm with two solutions indicated by $h^1(y)$ and  $h^2(y)$.  (b) Mapping by permutation operator $P$ indicated by the  gate ``Unitary'' of the register  $ |y  \rangle=\frac{1}{\sqrt{2}} \cdot ( |011 \rangle + |101  \rangle) $  into  $ |x  \rangle=|00 \rangle \otimes \frac{1}{\sqrt{2}} \cdot ( |0 \rangle + |1  \rangle)$. (c) Measuring the circuit (a). (d) Measuring the circuit (b) representing $P \cdot  \left(\frac{1}{\sqrt{2}} \cdot ( |011 \rangle + |101  \rangle) \right) =|00 \rangle \otimes \frac{1}{\sqrt{2}} \cdot ( |0 \rangle + |1  \rangle).
$ }
\label{D_el_v3_0.eps}
\end{figure}

\begin{figure}
\leavevmode
\parbox[b]{13cm}{ (a) \epsfxsize14cm\epsffile{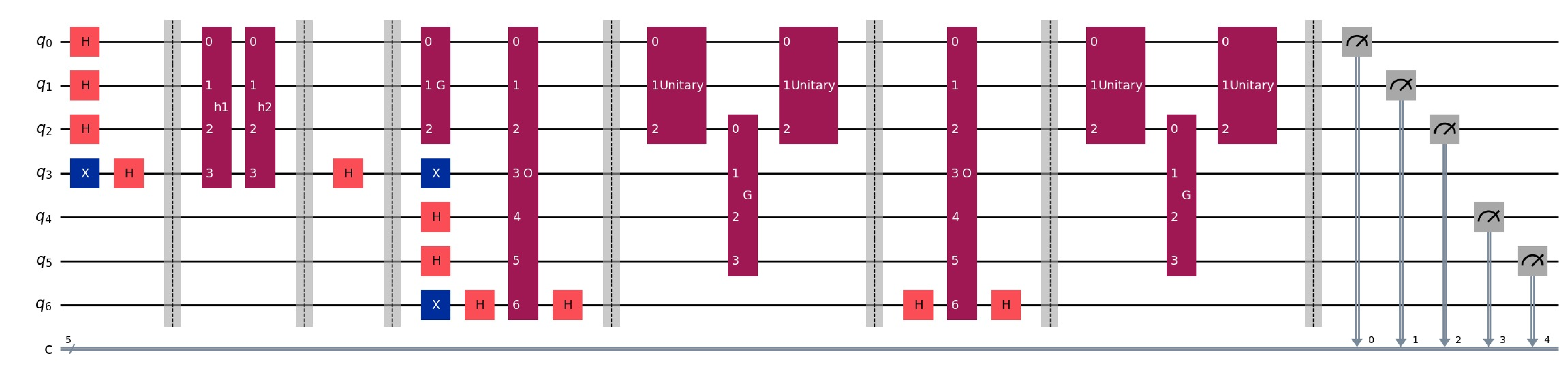}}
\begin{center}
\parbox[b]{10cm}{ (b) \epsfxsize9cm\epsffile{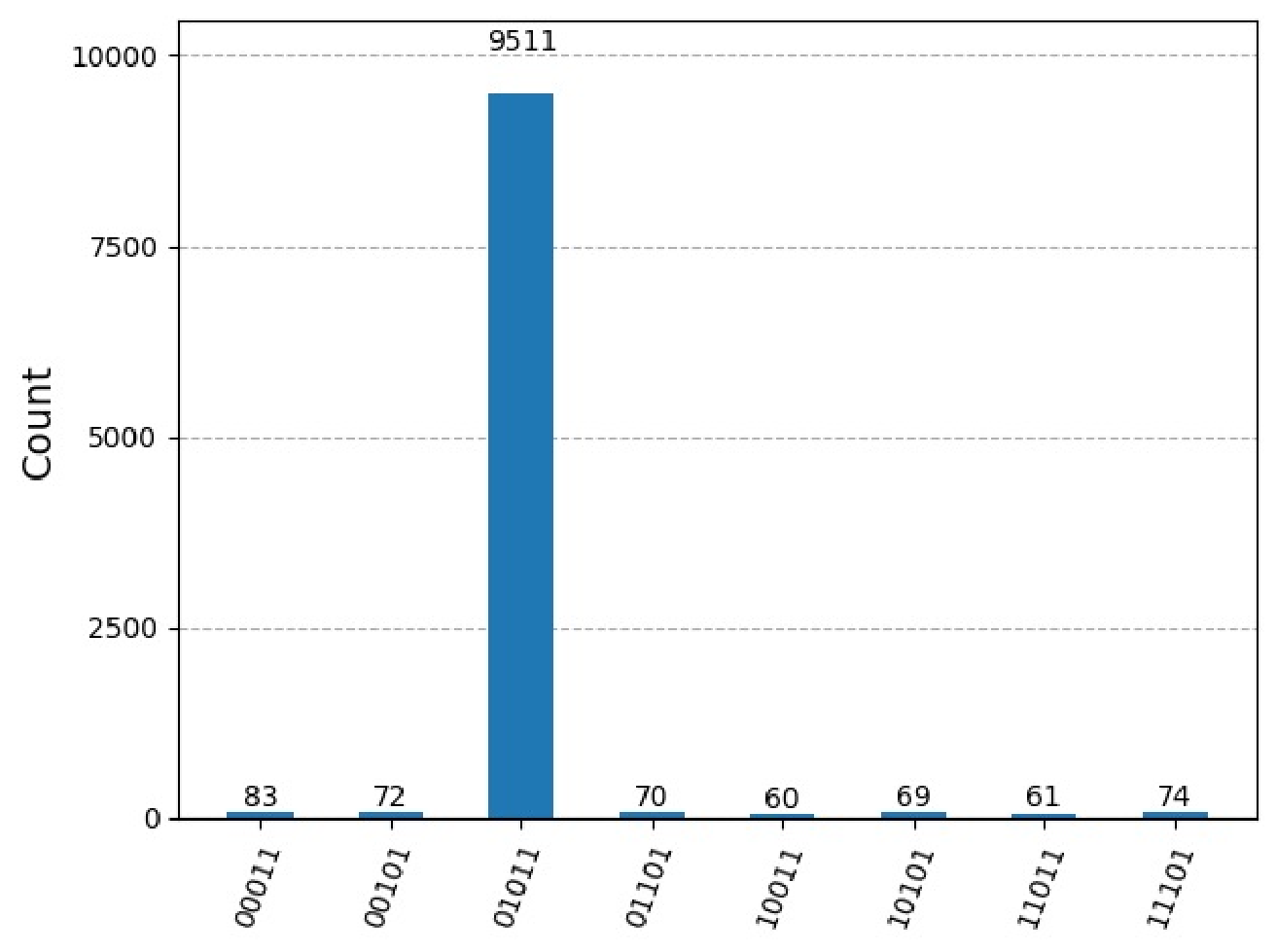}}
\end{center}
\caption{(a) The corresponding circuit, the permutation operator $P*$ is indicated by the  gate ``Unitary''. After applying Grover's algorithm with two solutions indicated by $h^1(y)$ and  $h^2(y)$.    using the oracle  on the original space $\frac{1}{\sqrt{2}} \cdot ( -|011 \rangle + |101  \rangle)$ to mark the solution, mapping  
$P^* \cdot  \left(\frac{1}{\sqrt{2}} \cdot ( -|011 \rangle + |101  \rangle) \right) =\frac{1}{\sqrt{2}} \cdot (- |0 \rangle + |1  \rangle)  \otimes  |00 \rangle$. 
Qubit  $2$ indicates the the marked solution describing the subspace $\mathcal{L}$ by permuted  values $h_1$ and $h_2$. Qubits $4$ and $5$ correspond to the subspace $\mathcal{U}$.   Grover's algorithm is applied on the  qubits $2$, $4$ and $5$.  We the transpose  $P^*$ to determine the solution by the oracle $o_{\xi}(x)$ in the original space. This is quite elegant, but until now, there appears to be no straightforward method to ascertain the mapping represented by $P^*$. (b) Determining the correct solution in the original space. }
\label{D_el_v_3_6.eps}
\end{figure}

We can map the set of solutions determined by the function of the set $\aleph$ into  in the subspace $\mathcal{L}$ into  another ordering that uses only the corresponding subset that could lead to a decomposition. For example the superposition
\[
\frac{1}{\sqrt{2}} \cdot ( |011 \rangle + |101  \rangle)
\]
is mapped  by a permutation matrix $P$
\[
P \cdot  \left(\frac{1}{\sqrt{2}} \cdot ( |011 \rangle + |101  \rangle) \right) = \frac{1}{\sqrt{2}} \cdot ( |000 \rangle + |001  \rangle) =
\]
\begin{equation}
 |00 \rangle \otimes \frac{1}{\sqrt{2}} \cdot ( |0 \rangle + |1  \rangle)
\end{equation}
into the decomposition $ |00 \rangle \otimes \frac{1}{\sqrt{2}} \cdot ( |0 \rangle + |1  \rangle)$, see Figure  \ref{D_el_v3_0.eps}.
We can represent  the permutation matrix $P$ as
\[
\left( \begin{array}{r} 
1\\
1\\
0\\
0\\
0\\
0\\
0\\
0\\
 \end{array} \right)=
 \left( \begin{array}{rrrrrrrr} 
0 & 0 & 0 & 1 & 0 & 0 & 0 & 0 \\
0 & 0 & 0 & 0 & 0 & 1 & 0 & 0 \\
0 & 0 & 1 & 0 & 0 & 0 & 0 & 0 \\
1 & 0 & 0 & 0 & 0 & 0 & 0 & 0\\
0 & 0 & 0 & 0 & 1 & 0 & 0 & 0\\
0 & 1 & 0 & 0 & 0 & 0 & 0 & 0\\
0 & 0 & 0 & 0 & 0 & 0 & 1 & 0\\
0 & 0 & 0 & 0 & 0 & 0 & 0 & 1\\
 \end{array} \right) \cdot
 \left( \begin{array}{r} 
0\\
0\\
0\\
1\\
0\\
1\\
0\\
0\\
 \end{array} \right).
\]
The  operation corresponds to the mapping by permutation operator $P$ of a register  $ |y_k  \rangle$ of $m$ qubits into a a register  $ |x_k  \rangle$ of $m$ qubits which first $g$ qubits are $ |0 \rangle ^{\otimes g}$ with
\begin{equation}
\frac{1}{\sqrt{v}}\sum_{k=1}^v P \cdot |y_k  \rangle  =  |0 \rangle ^{\otimes g} \otimes \left( \frac{1}{\sqrt{v}}\sum_{k=1}^v  |x_k \rangle \right). 
\end{equation}
Since Qiskit uses  little-endian notation we indicate  the inverse order  by a permutation matrix $P^*$
\[
P^* \cdot  \left(\frac{1}{\sqrt{2}} \cdot ( |011 \rangle + |101  \rangle) \right) = \frac{1}{\sqrt{2}} \cdot ( |000 \rangle + |100  \rangle) =
\]
\begin{equation}
\frac{1}{\sqrt{2}} \cdot ( |0 \rangle + |1  \rangle)  \otimes  |00 \rangle
\end{equation}
into the decomposition $ \frac{1}{\sqrt{2}} \cdot ( |0 \rangle + |1  \rangle)  \otimes  |00 \rangle$.
We can represent  the permutation matrix $P*$ due to  little-endian notation we get
\[
\left( \begin{array}{r} 
1\\
0\\
0\\
0\\
1\\
0\\
0\\
0\\
 \end{array} \right)=
 \left( \begin{array}{rrrrrrrr} 
0 & 0 & 0 & 1 & 0 & 0 & 0 & 0 \\
0 & 1 & 0 & 0 & 0 & 0 & 0 & 0 \\
0 & 0 & 1 & 0 & 0 & 0 & 0 & 0 \\
1 & 0 & 0 & 0 & 0 & 0 & 0 & 0\\
0 & 0 & 0 & 0 & 0 & 1& 0 & 0\\
0 & 0 & 0 & 0 & 1 & 0 & 0 & 0\\
0 & 0 & 0 & 0 & 0 & 0 & 1 & 0\\
0 & 0 & 0 & 0 & 0 & 0 & 0 & 1\\
 \end{array} \right) \cdot
 \left( \begin{array}{r} 
0\\
0\\
0\\
1\\
0\\
1\\
0\\
0\\
 \end{array} \right).
\]
The  operation corresponds to the mapping by permutation operator $P^*$ of a register  $ |y_k  \rangle$ of $m$ qubits into a a register  $ |x_k  \rangle$ of $m$ qubits which last $g$ qubits are $ |0 \rangle ^{\otimes g}$ with
\begin{equation}
\frac{1}{\sqrt{v}}\sum_{k=1}^v P^* \cdot |y_k  \rangle  =  \left( \frac{1}{\sqrt{v}}\sum_{k=1}^v  |x_k \rangle \right) \otimes  |0 \rangle ^{\otimes g}. 
\end{equation}
We can apply such a permutation operator $P^*$ for nested Grover’s algorithm for tree search as indicated  in the Figure \ref{D_el_v_3_6.eps}. The Grover's algorithm is applied on the  qubits $2$, $4$ and $5$.

\subsection{Costs}

Ignoring the costs of the permutation operator $P$ with the preparation costs $v$ by basis encoding in subspace $\mathcal{L}$, as described by \cite{Trugenberger2001,Trugenberger2003, Trugenberger2022}, the costs   corresponding to the Grover's algorithm applied to $m/2+\log_2 v$ qubits are
\begin{equation}
v +  \sqrt{2^{m/2+\log_2 v}} =v +  \sqrt{2^{m/2}} \cdot \sqrt{  v} . 
 \end{equation} 
With the preparation costs $\sqrt{ \frac{2^{m/2}}{v}}$ using Grover's algorithm in subspace  $\mathcal{L}$ the costs are
\begin{equation}
\sqrt{ \frac{2^{m/2}}{v}} +  \sqrt{2^{m/2+\log_2 v}}  =   \sqrt{2^{m/2}} \cdot \left( \frac{1}{\sqrt{ v} } + \sqrt{  v}  \right)
 \end{equation} 
 in $O$ notation given $v \approx m$ with
\[ O(v^{1/2} \cdot  n^{1/4})=O((\log n)^{1/2} \cdot  n^{1/4}).\]

\subsection{Permutation operator $P$ }

\begin{figure}
\leavevmode
\parbox[b]{13cm}{ (a) \epsfxsize14cm\epsffile{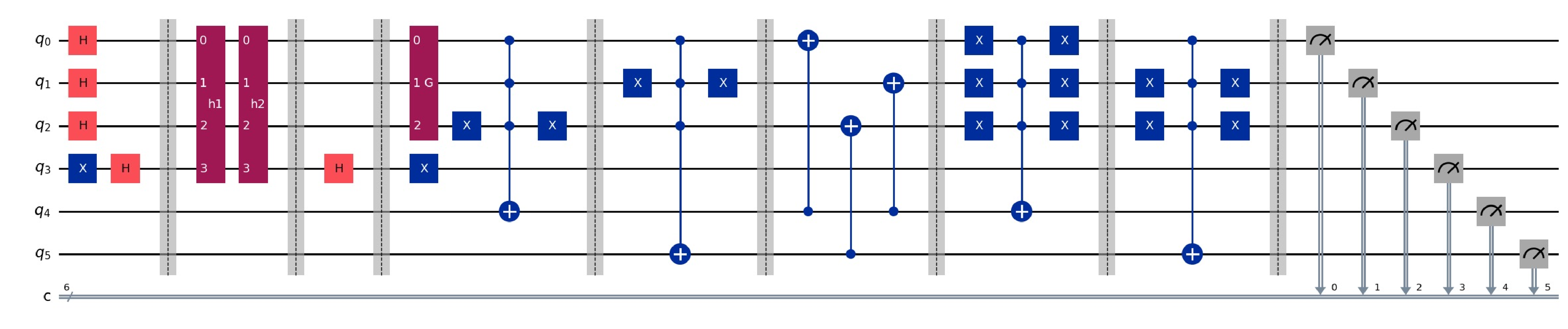}}
\begin{center}
\parbox[b]{7cm}{ (b) \epsfxsize6cm\epsffile{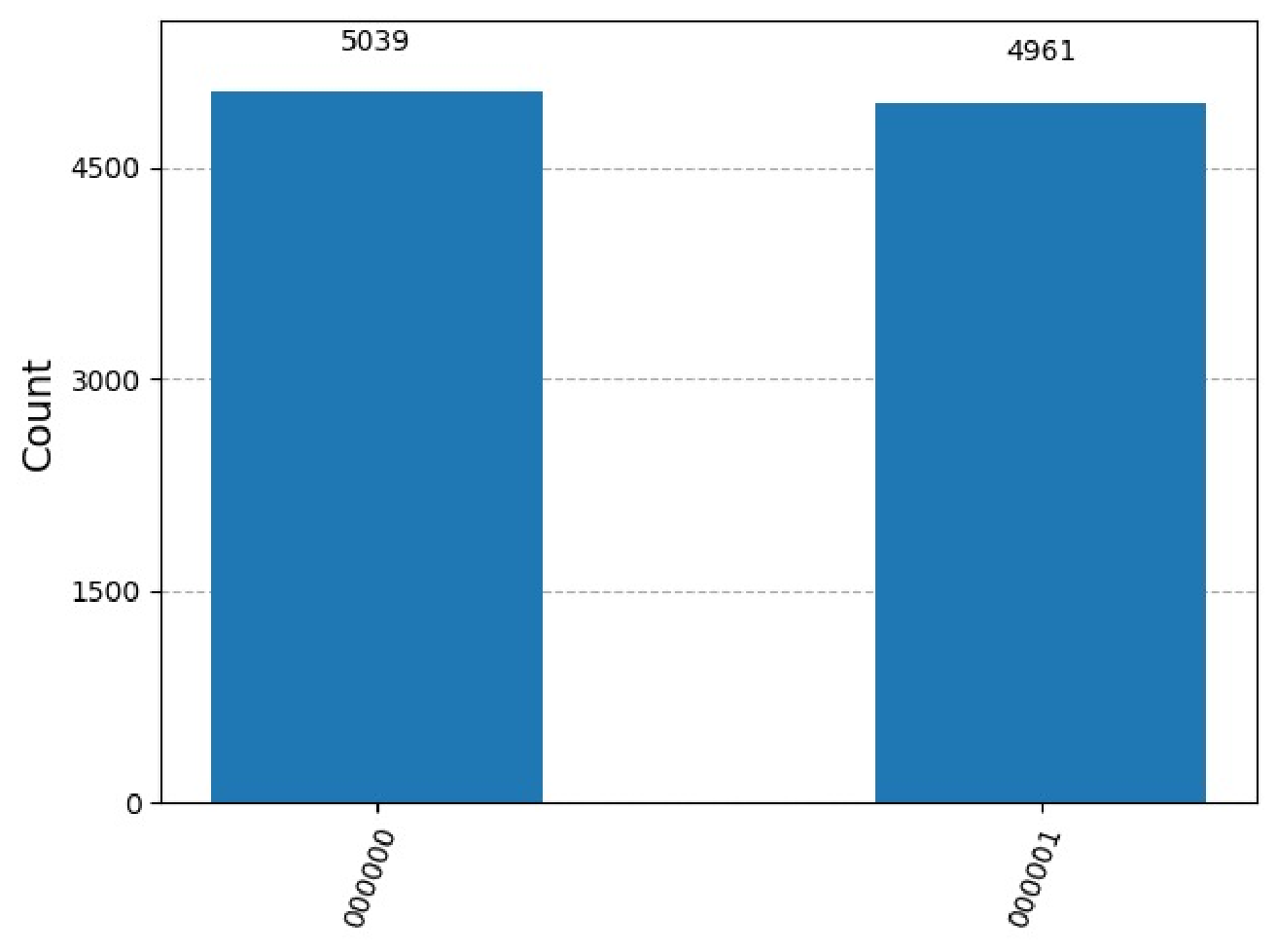}}
\parbox[b]{7cm}{ (c) \epsfxsize6cm\epsffile{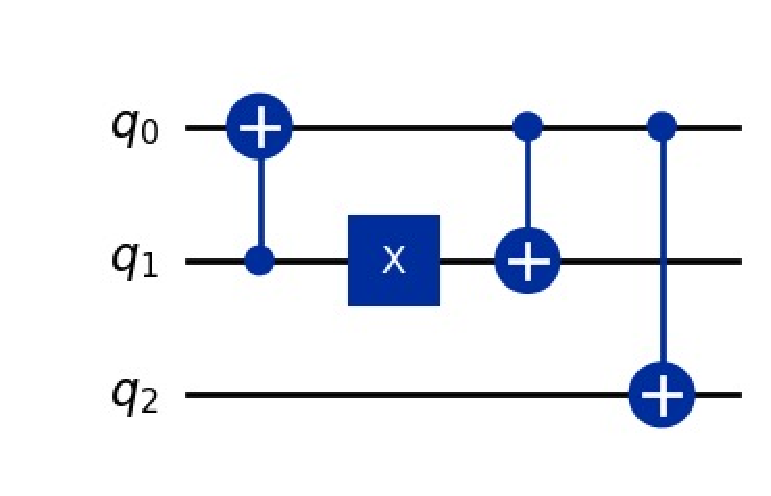}}
\end{center}
\caption{(a) Representation of the permutation operator $P \cdot  \left(\frac{1}{\sqrt{2}} \cdot ( |011 \rangle + |101  \rangle) \right) =|00 \rangle \otimes \frac{1}{\sqrt{2}} \cdot ( |0 \rangle + |1  \rangle)$  by controlled not gates. First we indicate the two paths $h_k$ in superposition by two flag qubits $4$ and $5$. We used the two flag qubits and the knowledge of the    $2$ paths $h_1, h_2$ to preform the permutation mapping. Knowing the permutation mapping we dis entangle the two  two flag qubits  $4$ and $5$ by mapping  them to zero. (b) The correct mapping $|00 \rangle \otimes \frac{1}{\sqrt{2}} \cdot ( |0 \rangle + |1  \rangle)$ without entanglement of the additional qubits $3$, $4$ and $5$ that are mapped to $0$. (c) A compressed representation of the permutation operator $P$.}
\label{E_el_4_4.eps}
\end{figure}

There appears to be no straightforward method to ascertain the mapping represented by $P$.
The  permutation operator $P$  is defined by the  $v$  paths $h_k$ to partial  could-be solutions. Due to entanglement we cannot determine the operator $P$ using   $h^k(z)$ and the resulting  $h_k$ paths. 
Due to entanglement we cannot determine the  permutation operator $P$ without knowing $h_k$ paths that indicate which nodes indicate partial  could-be solutions.
 Knowing  $h_k$ paths we can determine the permutation operator $P$ by  controlled not gates, see Figure \ref{E_el_4_4.eps} (a).  Using the regularities permutation operator $P$ can be considerably compressed, see Figure \ref{E_el_4_4.eps} (c) leading to reduced costs.
Knowing $h_k$ paths defines distinct nodes of the same depth simultaneously. For instance, in the quantum search tree depicted in Figure \ref{Tree_2_3_heuristics.eps}, the two extended oracles would be $o_1(011,u)$ and $o_2(000,u)$, with $g=3$, and we would seek the solution $u$.
This approach is appropriate for a tree search where we presume that certain nodes at a depth $g$ signify potential solutions. Alternatively, we could implement the two extended oracles directly through a quantum circuit using concatenation.

\section{Discussion}

\subsection{Partial Candidate Solution $h(y)$}

Within the context of quantum tree search, instead of employing a heuristic function $\mu(y)$, we will utilize the function $h(y)$, which signifies a potential solution and $y$ represents a path from the root to the corresponding node. In contrast to the heuristic function $\mu(y)$, which can mark multiple nodes, $h(y)$ should exclusively mark a single potential solution for the node.
A set of partial candidate solutions can be identified by determining their similarity or distance to the goal state.
However, how can we pinpoint a single potential solution, a single partial candidate solution?
One approach to achieve this is to introduce an additional constraint that narrows the set to a single function, denoted as $h_k(y)$. Such a constraint could, for instance, specify the possible position within a subtree, thereby reducing the set to a single function. It is important to note that this approach is highly probabilistic and may occasionally yield non-existent partial candidate solutions or values exceeding one. Additionally we have to take care that $v$ is constraint, as indicated by the Equation \ref{v:const}.
Additional in further research has to investigate partial candidate solution $h_k(y)$ function on examples as indicated in the paper \cite{Wichert2022b}.

\subsection{Generalized Quantum Tree Search}
  
 We must have prior knowledge of the depth $m$ off the search tree. This constraint can be circumvented through iterative deepening. In iterative deepening search, we progressively increase the search limit from one to two to three to four and continue until a goal is discovered. For each limit, a search is conducted from the root to the maximum depth of the search tree. Should the search prove unsuccessful, a new search is initiated with a deeper limit.
During iterative deepening search, states are generated multiple times \cite{korf1985}, \cite{russell2010}.
The time complexity of iterative deepening search is comparable to that of a search to the maximum depth \cite{korf1985}. A quantum iterative deepening search is equivalent to iterative deepening search \cite{tarrataca2012b}, \cite{Tarrataca2013}.
For each limit $max$, a quantum tree search is performed from the root, where $max$ represents the maximum depth of the search tree. The potential solutions are determined through measurement. 

For a not constant branching factor the quantum tree search the maximal branching factor $B_{max}$ has to be used for the quantum tree search   \cite{tarrataca2010}. For the maximum value of $B_{max}$, the quantum algorithm using qubit representation outperforms the classical tree search described by the effective branching factor $b$ in the specified case
\begin{equation}
b > b_q = \sqrt{B_{max}}.  
 \end{equation} 
For a large number of instances with varying initial and goal states, the effective branching factor converges to the average branching factor for an uninformed tree search, as shown in \cite{tarrataca2010}.

Consider each branching factor as a potential local path.
For a non-constant branching factor, the quantum tree search utilizes the maximal branching factor, denoted as $B_{max}$. If, in a node $\nu$, $B_{\nu} < B_{max}$, only $B_{\nu}$ are required, and $B_{max}-B_{\nu}$ local paths are not utilized. To address this limitation, we augment the path descriptor by repeating the paths. This approach ultimately leads to $k$ solutions that converge to the effective branching factor, as demonstrated in \cite{Wichert2022b}, \cite{Wichert2024}.

\section{Conclusion}

Conventional heuristic functions that cannot be applied to quantum tree search. 
 If we overcome the limitations of the definition and existence of the  partial candidate solution  $h(y)$, then  the iterative approach represents an considerably speed up with the   costs  
\begin{equation}
 v \cdot \left(\sqrt{2^{m/2}} + \sqrt{2^{{m/2}}} + 1 \right) =   v \cdot \left(2 \cdot  2^{{m/4}} +1  \right)
 \end{equation} 
We can express the cost of iterative search it in $O$ notation given $v \approx m$ since $n=2^m$ as 
\[ O(v \cdot  n^{1/4})=O(\log n \cdot  n^{1/4}).\]

Combinatory approach with possible existence of  the upper oracle $u(z)$ 
allows us to  disentanglement of $\mathcal{L}$ and $\mathcal{U}$ using 
\[
\mathcal{H}=\mathcal{U}_{ h_k \neq h_1}    \otimes \mathcal{U}_{ h_k \neq h_2}   \otimes \cdots  \otimes \mathcal{U}_{ h_k \neq h_v}   \otimes \mathcal{L}
\]
speeds up the costs minimally.
Finally, we presented the theoretical concepts of the potential permutations within the subspace $\mathcal{L}$. The work suggests a novel direction based on a partial candidate solution rather than heuristic functions that are unsuitable for quantum tree search.  All Qiskit examples are presented in a Jupyter Notebooks that can be freely downloaded at https://github.com/andrzejwichert/Nested-Quantum-Tree-Search.

\section{Compliance with Ethical Standards}

The funders had no role in study design, data collection and analysis, decision to publish, or preparation of the manuscript. The authors declare no conflicts of interest. This article does not contain any studies with human participants or animals performed by any of the authors.

\bibliographystyle{plainnat}

\end{document}